\documentclass[journal]{IEEEtran}

\ifCLASSINFOpdf
\else
   \usepackage[dvips]{graphicx}
\fi
\usepackage{url}
\usepackage{color}
\usepackage{stfloats}

\usepackage{graphicx}

\begin{document}

\title{Superpixel Tensor Pooling for Visual Tracking using Multiple Midlevel Visual Cues Fusion}

\author{Chong Wu, \IEEEmembership{Student Member, IEEE}, Le Zhang, Jiawang Cao, and Hong Yan, \IEEEmembership{Fellow, IEEE}
\thanks{This article has been accepted for publication in a future issue of IEEE Access, \emph{Digital Object Identifier: 10.1109/ACCESS.2019.2946939}, you can cite as: C. Wu, L. Zhang, J. Cao and H. Yan, ``Superpixel Tensor Pooling for Visual Tracking using Multiple Midlevel Visual Cues Fusion," in \emph{IEEE Access}. DOI: 10.1109/ACCESS.2019.2946939.
}
\thanks{Chong Wu and Hong Yan are with the Department of Electrical Engineering, City University of Hong Kong, Kowloon, Hong Kong (e-mail: chongwu2-c@my.cityu.edu.hk \& h.yan@cityu.edu.hk).}
\thanks{Le Zhang is with the Department of Computer Science and Technology, Tongji University, Shanghai 200092, China (e-mail: zhangle\_tj@163.com).}
\thanks{Jiawang Cao is with the School of Automation, China University of Geosciences, Wuhan 430074, China  (e-mail: CJW@cug.edu.cn).}}

\markboth{}
{}
\maketitle

\begin{abstract}
In this paper, we propose a method called superpixel tensor pooling tracker which can fuse multiple midlevel cues captured by superpixels into sparse pooled tensor features. Our method first adopts the superpixel method to generate different patches (superpixels) from the target template or candidates. Then for each superpixel, it encodes different midlevel cues including HSI color, RGB color, and spatial coordinates into a histogram matrix to construct a new feature space. Next, these matrices are formed to a third order tensor. After that, the tensor is pooled into the sparse representation. Then the incremental positive and negative subspaces learning is performed. Our method has both good characteristics of midlevel cues and sparse representation hence is more robust to large appearance variations and can capture compact and informative appearance of the target object. To validate the proposed method, we compare it with state-of-the-art methods on 24 sequences with multiple visual tracking challenges. Experiment results demonstrate that our method outperforms them significantly.
\end{abstract}

\IEEEpeerreviewmaketitle

\section{Introduction}
\IEEEPARstart{T}{he} study of visual tracking has been achieved great successes in recent years. Visual tracking is a process of locating a moving object or multiple objects over time in a video stream or using a camera. It can be divided into three steps: (1) object detection; (2) location prediction; (3) data association. Before using tracking algorithm to perform these steps, for each video application, a shot boundary detection needs to be performed to extract the sequence \cite{SBD}. However, because of the heavy occlusion, drifts, fast motion, severe scale variation, large shape deformation, \emph{etc.}, visual tracking is still a challenge in computer vision \cite{SPL1, TPT2, TPT}.

Many advanced visual tracking methods have been developed to solve these challenges, such as sparse representation based approaches, correlation filter (CF) based methods, deep learning (DL) based methods, \emph{etc}. Sparse representation has been introduced successfully into the construction of the appearance model in visual tracking \cite{TPT2, TPT, INCLF}. It uses the sparse linear representation to represent the candidates \cite{TPT2, INCLF}. It can use very few but most related target templates to reduce impacts of background noise \cite{TPT}. Moreover, it can use local sparse codes to model the target appearance adaptively and exploit the discriminative nature \cite{TPT}. Some effective sparse representation based methods has been proposed \cite{TPT2, TPT, INCLF}. The work in \cite{TPT} has introduced the tensor pooling into the construction of the discriminative appearance model, which can deliver intrinsic structural information and robust against drifting and environmental noise. Nonnegative local coordinate factorization in visual tracking and the inverse nonnegative local coordinate factorization into constructing the discriminative appearance model proposed in \cite{INCLF} shows a strong discriminative ability between the target and the background. However, most of these methods have a defect of high computational cost. CF based methods use the correlation filter theory from the signal processing field \cite{IBACF}. CF based methods can achieve a high-speed and real-time tracking and a high accuracy. But most of them does not work well in tracking a rapidly moving object, a large object scale variation, or a non-rigid object. To solve these problems, some improved CF based methods have been proposed \cite{BACF, KCFS, IBACF}. Recently, deep learning has been used successfully in the visual tracking \cite{KCFS}. Compared to conventional tracking methods, the DL based methods have shown larger improvements in tracking performance. The work in \cite{DLT} has introduced a convolutional neural network (CNN) into visual tracking, which is one of the first works of introducing CNN into the visual tracking. Another classical DL based trackers is the multi-domain convolutional neural networks (MDNet) \cite{MDNet}. It trains a multi-domain CNN to distinguish the object and background and has achieved high tracking accuracy and success rate \cite{BACF}. However, it suffers from a slow speed. To avoid the training problem, some works using pre-trained CNN for feature extraction have been developed. Paper \cite{HCF} uses a pre-trained CNN to obtain hierarchical convolutional features and improve accuracy and robustness.

One important aspect of visual tracking is the appearance modeling. Different levels of appearance and spatial cues are successfully applied in the appearance modeling \cite{SPT, SPT1}. Compared to high-level information (eg. structural information) and low-level visual cues (eg. pixels), midlevel visual cues (eg. group of pixels such as superpixels) are shown to be more effective in representing the structure of the image \cite{SPT, SPT1, ISPT, CST}. Some researchers applied superpixel methods to obtain the midlevel cues in visual tracking and their methods show robust against heavy occlusion and drifts \cite{SPT, SPT1, ISPT, CST, DSP}. Superpixel is a group of pixels which are similar in some properties like color \cite{SLIC, SNIC, FSLIC}. Superpixels can reduce the redundancy and preserve the structure from the image \cite{SLIC,FSLIC,ICIP1,JFast,FuzzyS}. Through substituting thousands of pixels with only hundreds of superpixels, subsequent image processing tasks are also speeded up \cite{SLIC,FSLIC,ICIP1,Harmony,SEEDS}. But the utilization of superpixels will reduce the dimension of original data (for a matrix, it will be vectorized). It will cause the loss of spatial information of the target object. Hence, how to fuse spatial information into the appearance model constructed by midlevel cues is still a challenge \cite{DSP}. Some researchers used the Euclidean distances from the target to the candidates as the weight to integrate spatial information \cite{SPT, SPT1}. It can preserve some spatial information to some extent. However, the spatial information like the shape of the target is still lost. Hence, some of these methods are more sensitive to color variations than spatial variations and this might result in a poor performance in a tracking challenge such as background clutters. Some researchers have shown that integrating more information in sparse representation can improve the tracking performance \cite{SPARSE}. Some other researchers tried to fuse the depth cue with superpixel-based target estimation using graph-regularized sparse coding and improved the discriminative ability of the trackers \cite{DSP}. Moreover, different color channels are suitable to different tracking scenarios. Integrating different color channels into a unified framework will also help to improve the robustness of tracker. Hence, it is of great interest to develop an elegant method for fusing multiple midlevel cues in sparse representation.
 \begin{figure*}
\centering 
\centerline{\includegraphics[width=17cm]{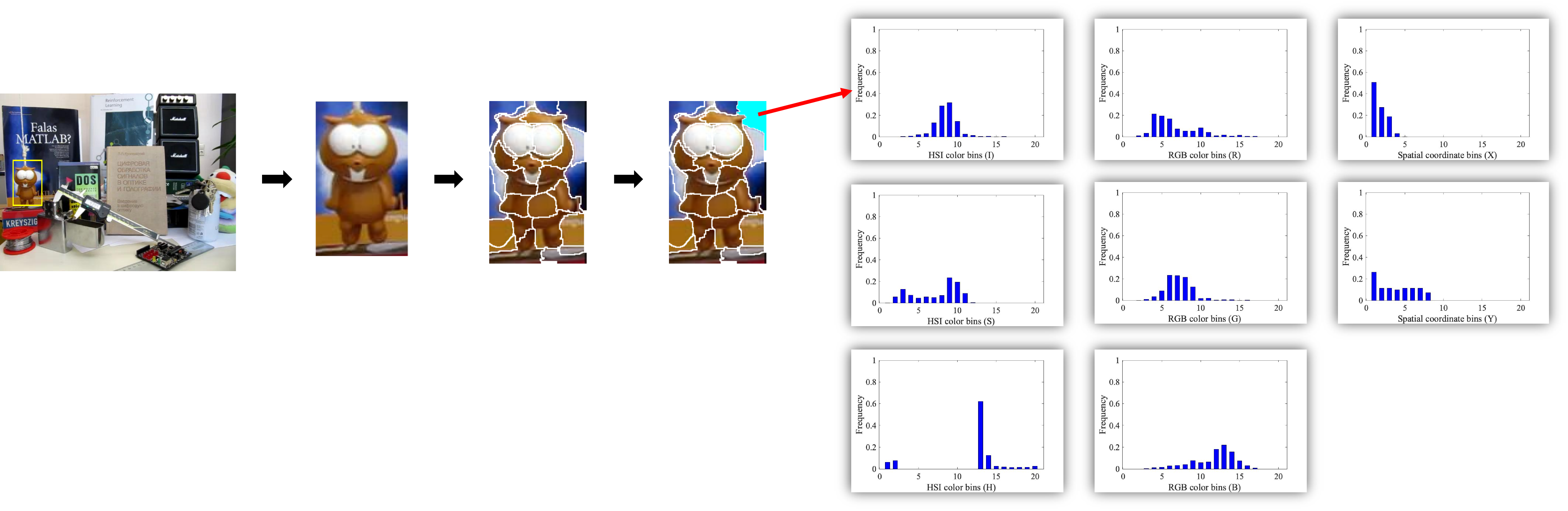}}
\caption{The process of patches extraction and features construction using the superpixel segmentation. (1) Select the target template of size $n_1*n_2$; (2) Generate $n_3$ superpixels to cover the template exhaustively; (3) Calculate the normalized histograms of multiple midlevel visual cues as the features for each superpixel.}
\label{fig0}
\end{figure*}

In this paper, we propose a visual tracking method called superpixel tensor pooling tracker (SPTPT) which can integrate multiple midlevel cues (such as the information of different color channels, spatial coordinates, shape, \emph{etc.}) obtained by superpixels in a unified sparse coding tensor form. With the utilization of sparse representation and midlevel cues, our method has both good characteristics of the midlevel cues and sparse representation. In addition, through fusing multiple midlevel cues, our method is more robust than some state-of-the-art methods under large appearance variations. The contribution of this paper can be summarized as follows,

\begin{itemize}
  \item [1)] 
  This is the first attempt of using superpixels to obtain tensor-pooled sparse features. With the utilization of superpixels, the patches obtained are more meaningful than the patches obtained by sliding window.
  \item [2)]
  Our method provides an effective fusion framework for fusing multiple midlevel cues in a unified sparse representation. Hence the constructed discriminative appearance model can take the advantage of different midlevel cues.
  \end{itemize}

To validate our method, we select 11 state-of-the-art tracking methods and 24 sequences with multiple tracking challenges from the benchmark \cite{Bench}, and we compare their reasonable lower-bound performances (we used one default parameter for all sequences without any tuning). Experiment results show that the lower-bound performance of our method is significantly better than existing ones.
 
The rest of this paper is organized as follows. In Section 2, we first introduce the superpixel segmentation method used in our method. Then we describe the fusion model and the incremental positive and negative subspaces learning method. At the end of this section, we introduce the motion model and give a brief summary of the proposed algorithm. In Section 3, we first illustrate the experiment settings and evaluation metrics. We then analyze the experiment results. Finally, we present the conclusion in Section 4.

\section{Superpixel tensor pooling tracker}
In this section, we will introduce four main parts and provide an algorithm summary of the proposed SPTPT.

\subsection{Patches Extraction using Superpixel Segmentation}
Producing meaningful patches is important to construct tensor-pooled sparse features. Compared to the patches obtained by sliding window \cite{TPT2, TPT}, superpixels are more meaningful, because superpixels can preserve the image structure and reduce the redundancy. Hence, we introduce superpixel segmentation into patches (superpixels) extraction in SPTPT. To construct tensor-pooled sparse features, we need to keep the number of patches obtained precisely. Hence, a superpixel method which can control the superpixel number precisely is needed. We select simple non-iterative clustering (SNIC) \cite{SNIC}, which can generate precise number of superpixels as the superpixel method in SPTPT (compactness coefficient: 20, in this paper). SPTPT uses a template of size $n_1*n_2$ to select a target template. Then, SNIC is adopted to generate $n_3$ non-overlapping superpixels to cover the template exhaustively. In this paper, we set $n_1 = n_2 = 32, n3 = 30$. We use normalized histograms in HSI color space, RGB color space, and spatial information respectively as the features for each superpixel. Each histogram vector for each midlevel cue in a superpixel is as follow,
\begin{equation}
\emph{\textbf{f}} = [f_1, f_2,...,f_n]^\mathrm{T},
\label{fun01}
\end{equation}
where, $n$ is the number of bins of each histogram. Each element in $\emph{\textbf{f}}$ represents the frequency of each bin in a superpixel region. It can be calculated as follow,
\begin{equation}
f_i = \frac{c}{r},
\label{fun02}
\end{equation}
where, $c$ is the number of pixels corresponding to bin $i$ in a superpixel region and $r$ is the total number of pixels in this superpixel region.

\subsection{Fusion Model for Multiple Midlevel Cues}
Before combining these vectors into a unified matrix form, it is important to evaluate the correlation between the HSI color space and the RGB color space. To calculate the correlation of different color spaces, we vectorize all images of the dataset we used in HSI color space and RGB color space respectively. Then we use the Pearson's linear correlation coefficient to calculate the correlation of these vectors pairwise. And finally, we get the overall Pearson's linear correlation coefficient $\rho = 0.0591$ and a p-value $p = 0.003 < 0.05$, which means they are correlated but the correlation is very weak. Hence, we can combine these color spaces directly. These vectors are combined to construct the feature matrix of each superpixel as follow,
\begin{equation}
\emph{\textbf{F}} = [\emph{\textbf{f}}_1,\emph{\textbf{f}}_2,...,\emph{\textbf{f}}_m],
\label{fun03}
\end{equation}
where, $m$ is the number of features, and in this paper, $m = 8$ (H, S, I, R, G, B, \emph{x}, \emph{y}). The process of patches extraction and features construction is shown in Fig. \ref{fig0}.
\begin{figure*}
\centering 
\centerline{\includegraphics[width=17cm]{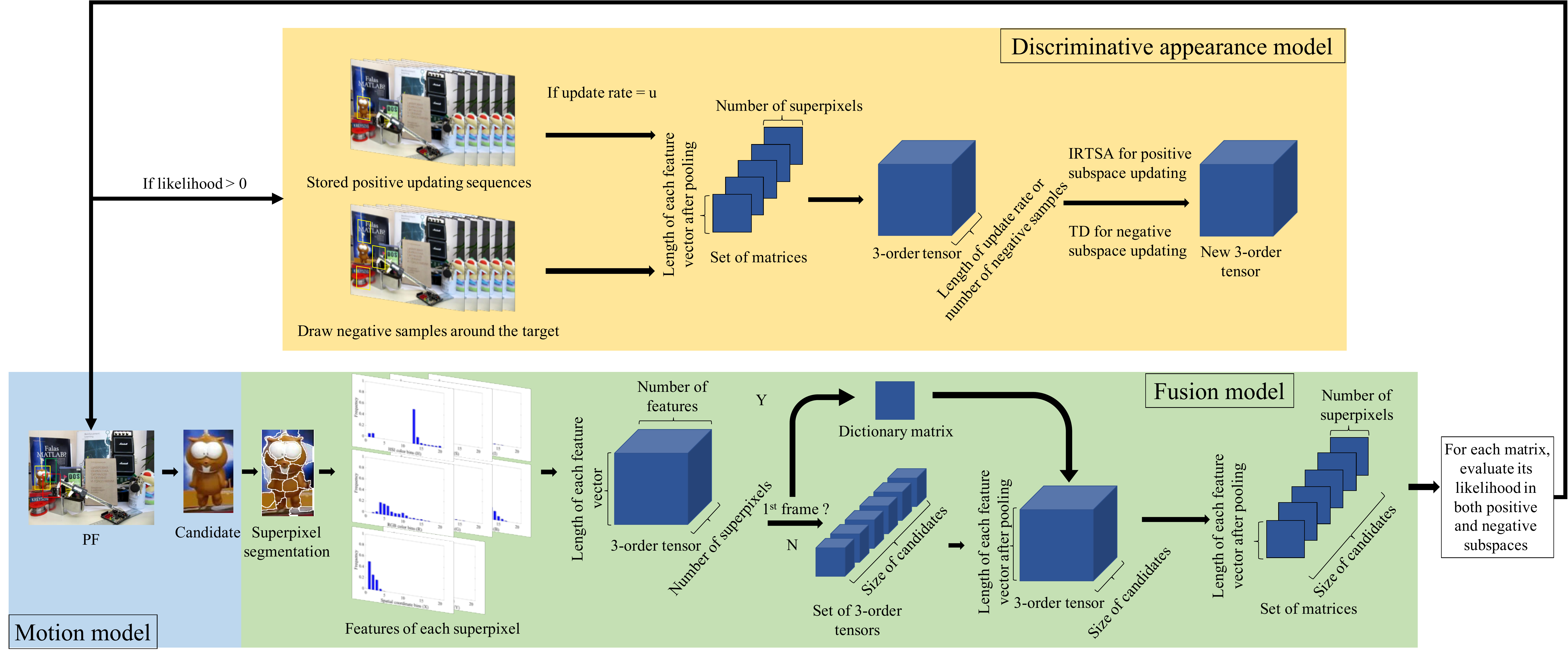}}
\caption{The workflow of the proposed tracker.}
\label{fig1}
\end{figure*}

We apply local sparse codes to encode these histogram features. We reshape the feature matrix to a feature vector $\emph{\textbf{a}}\in \textbf{R}^{B}$ of each superpixel and compute the sparse coefficient vector $\emph{\textbf{h}} \in \textbf{R}^z$ of it using the formula as follow,
\begin{equation}
\min\limits_{h_i} ||a_i - \emph{\textbf{D}}h_i||_2^2+\lambda||h_i||_1,
\label{fun04}
\end{equation}
where, $B = n \times m$, $\emph{\textbf{D}}\in\textbf{R}^{z\times s}$ is the dictionary matrix learned by the clustering result of $\emph{\textbf{a}}$ of the superpixels obtained in the 1st frame, $z$  is the number of cluster centroids and $s$ is the number of superpixels.

Then, we arrange the sparse coefficient vector $\emph{\textbf{h}}$ of each superpixel of each candidate in a unified 3-order tensor $\mathcal{T} \in \textbf{R}^{z \times s \times v}$ according to the spatial order of their corresponding superpixels in the candidate templates. $v$ is the number of candidate templates.
\begin{figure*}[h]
\centering
\begin{minipage}[b]{0.2\linewidth}
  \centering
  \centerline{\includegraphics[width=\linewidth]{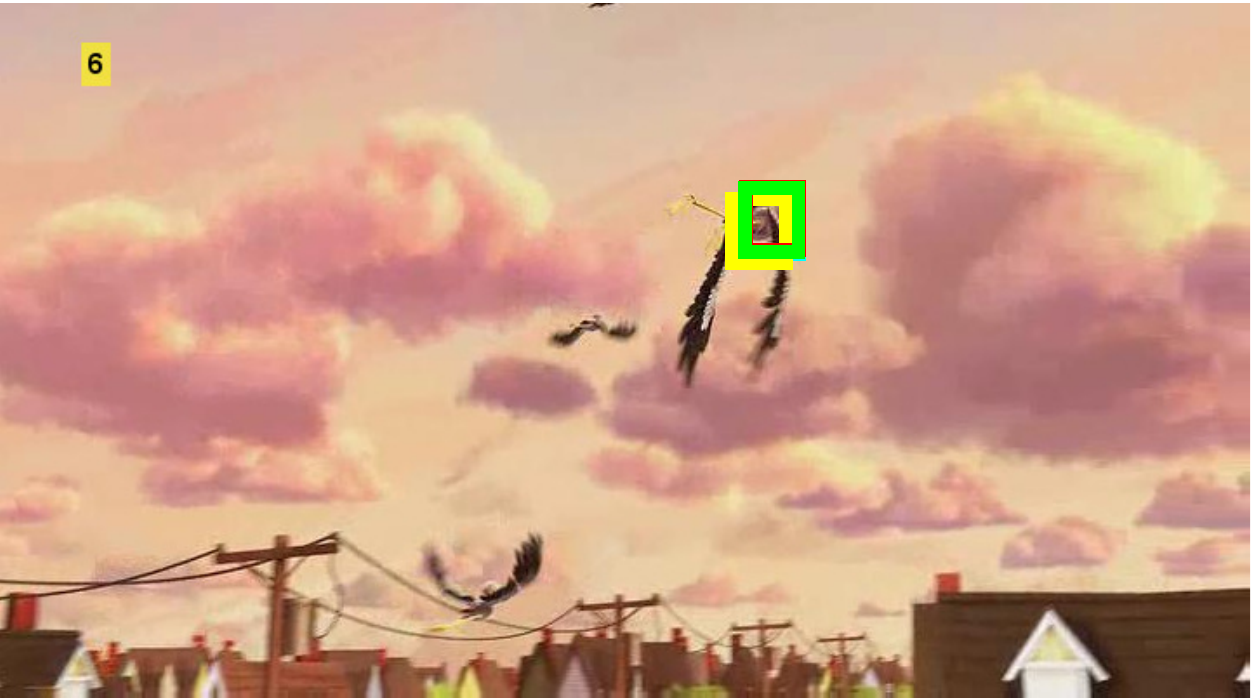}}\vspace{4.pt}
\end{minipage}
\begin{minipage}[b]{0.2\linewidth}
  \centering
  \centerline{\includegraphics[width=\linewidth]{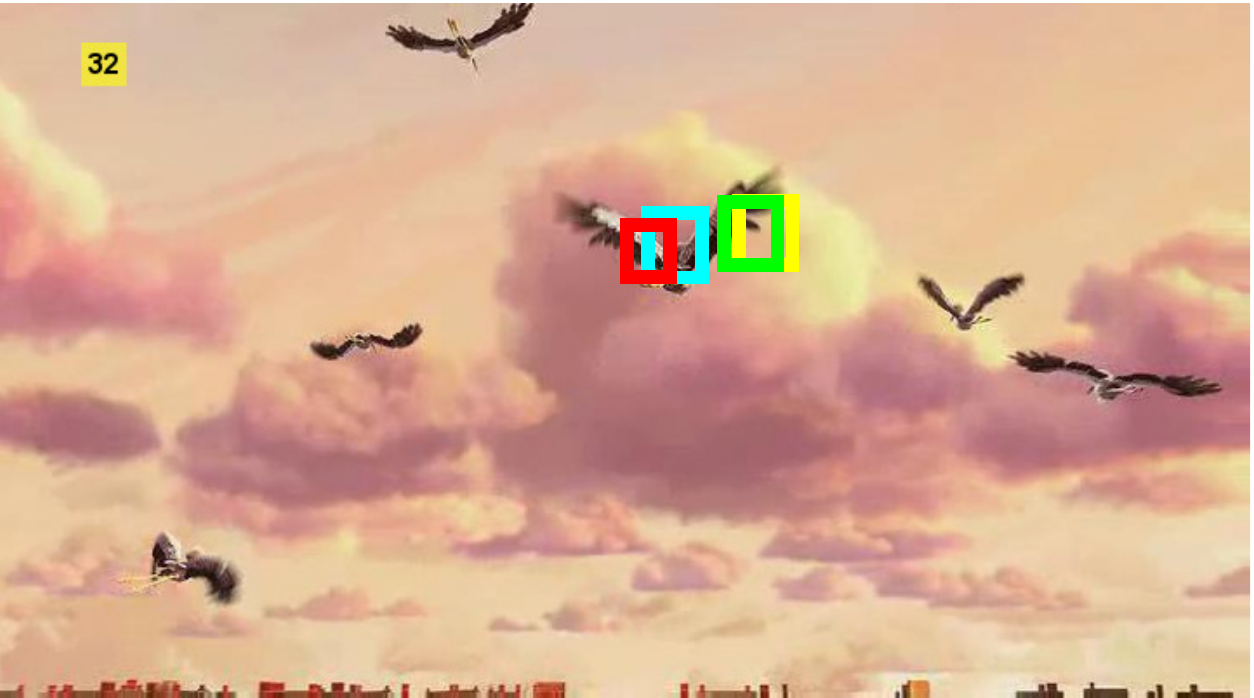}}\vspace{4.pt}
\end{minipage}
\begin{minipage}[b]{0.2\linewidth}
  \centering
  \centerline{\includegraphics[width=\linewidth]{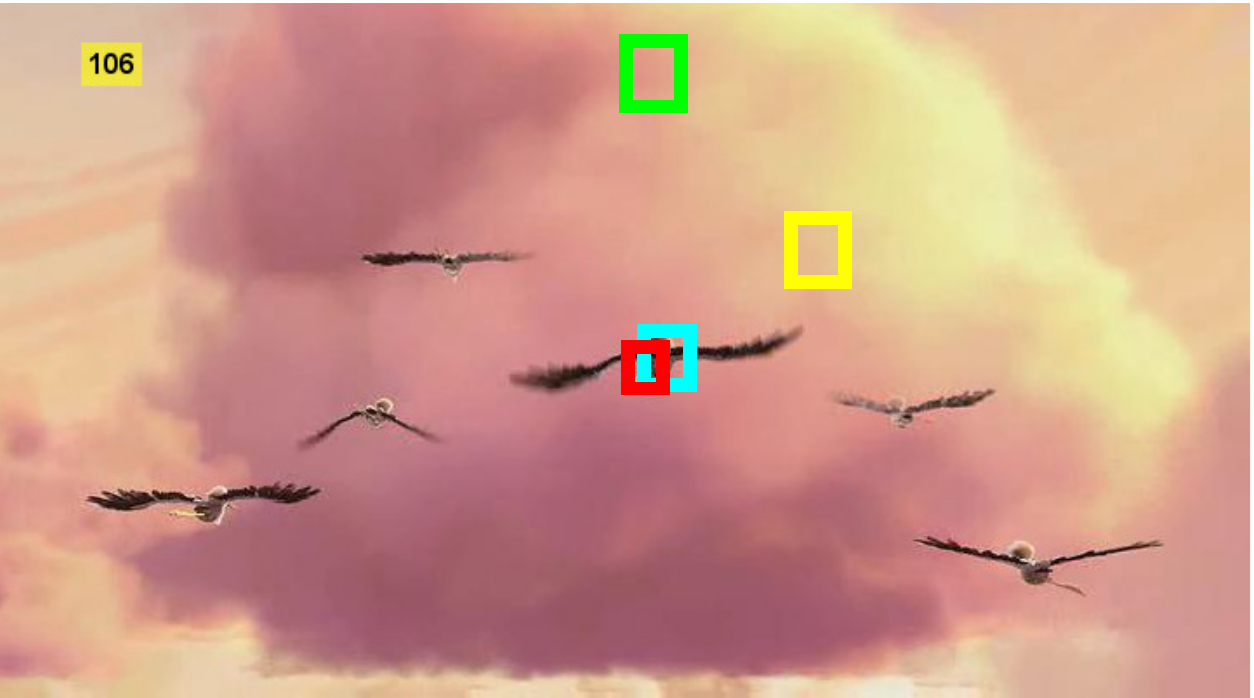}}\vspace{4.pt}
\end{minipage}
\begin{minipage}[b]{0.2\linewidth}
  \centering
  \centerline{\includegraphics[width=\linewidth]{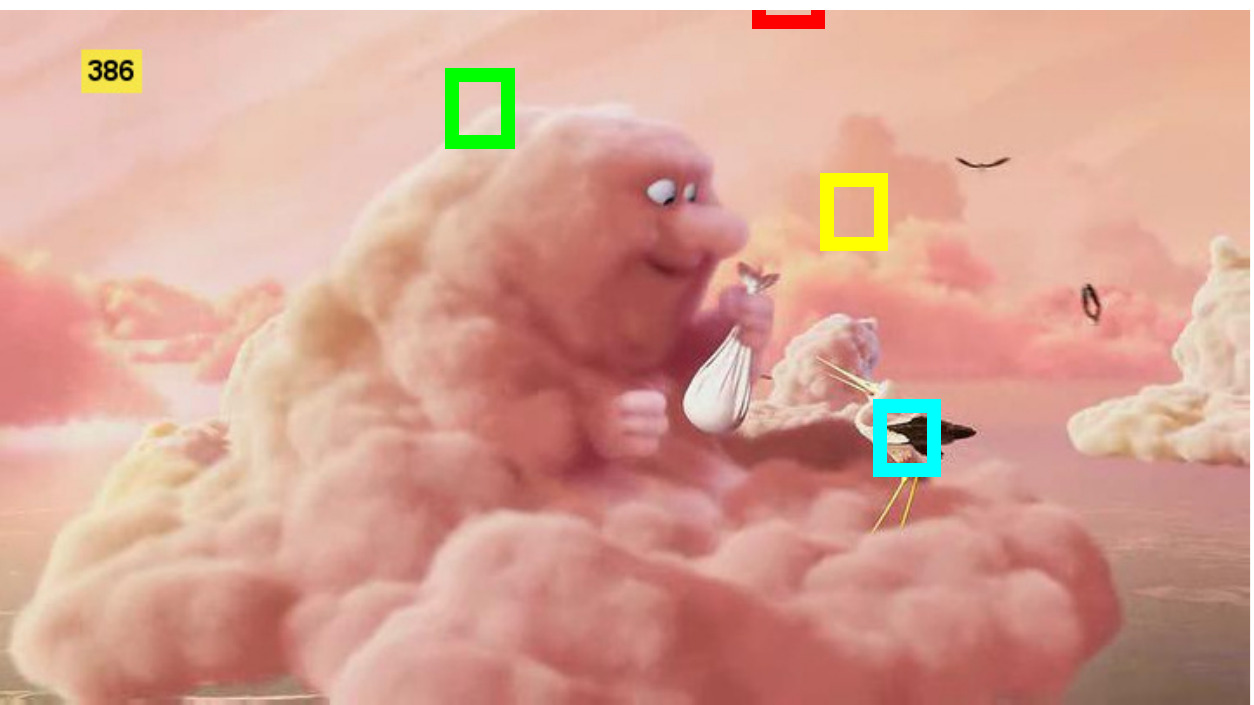}}  \vspace{4.pt}
  \end{minipage}

  \begin{minipage}[b]{0.2\linewidth}
  \centering
  \centerline{\includegraphics[width=\linewidth]{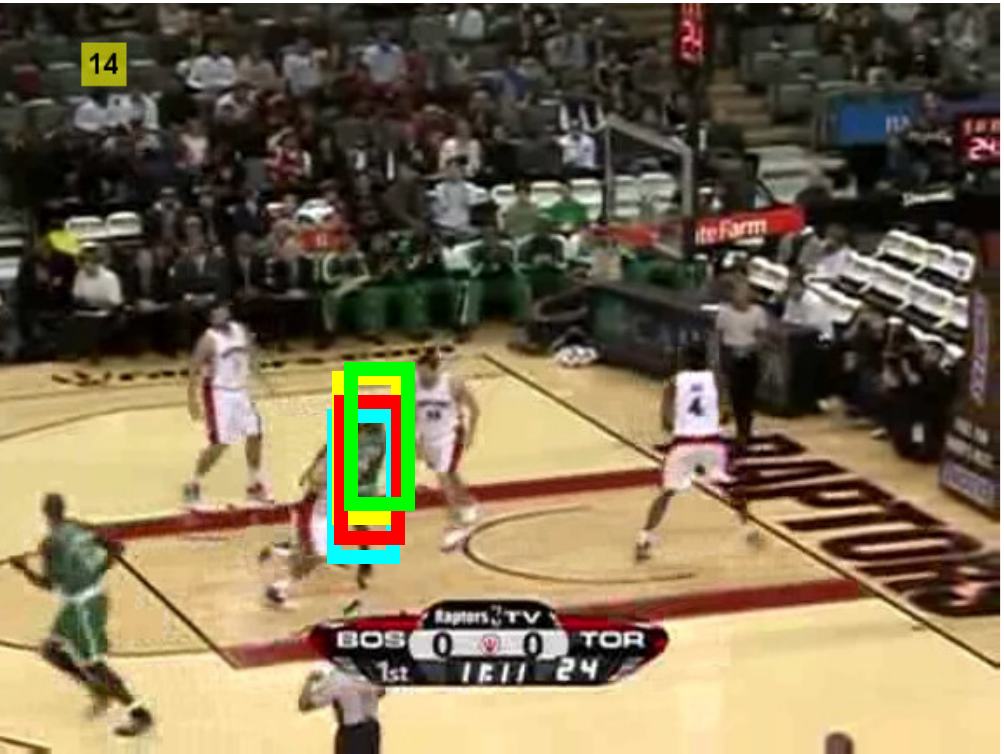}}\vspace{4.pt}
\end{minipage}
\begin{minipage}[b]{0.2\linewidth}
  \centering
  \centerline{\includegraphics[width=\linewidth]{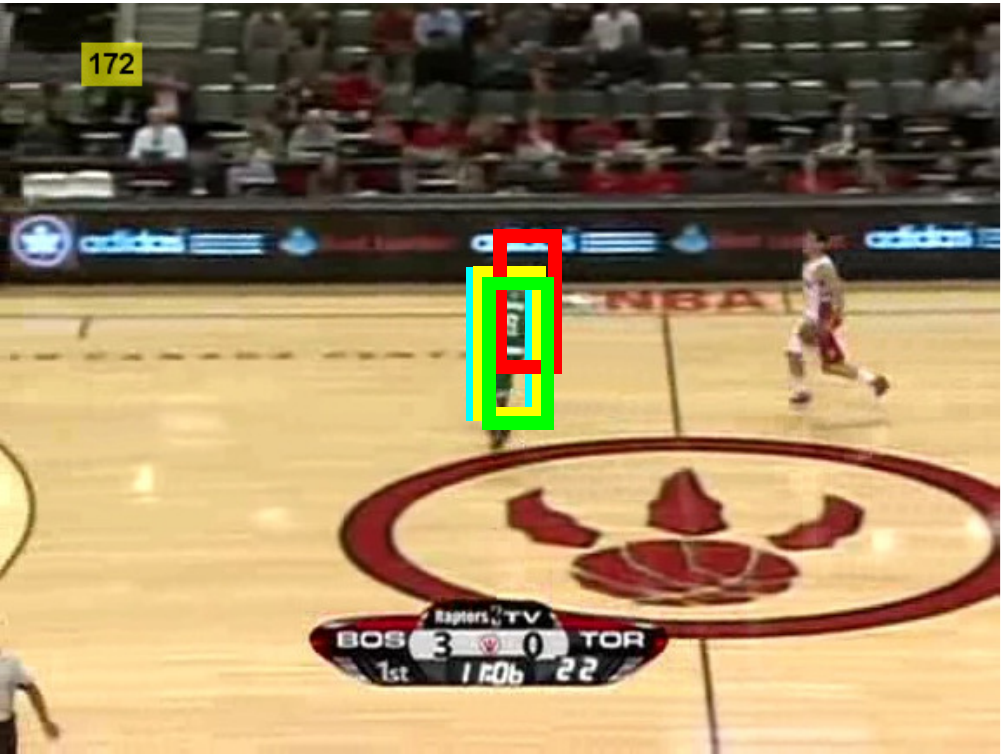}}\vspace{4.pt}
\end{minipage}
\begin{minipage}[b]{0.2\linewidth}
  \centering
  \centerline{\includegraphics[width=\linewidth]{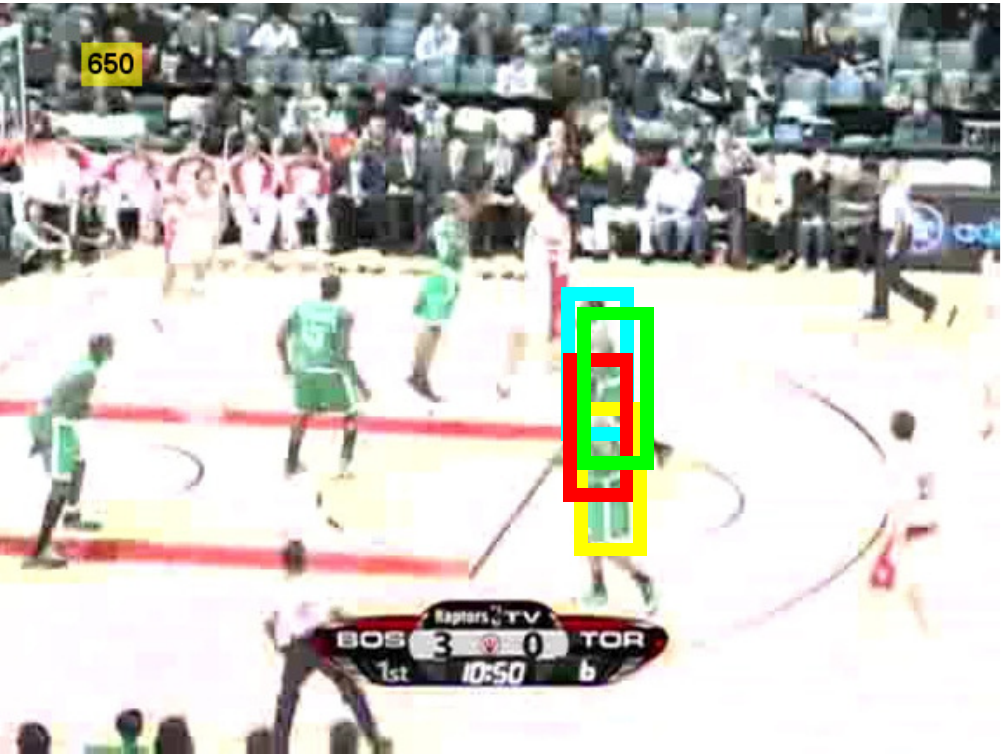}}\vspace{4.pt}
\end{minipage}
\begin{minipage}[b]{0.2\linewidth}
  \centering
  \centerline{\includegraphics[width=\linewidth]{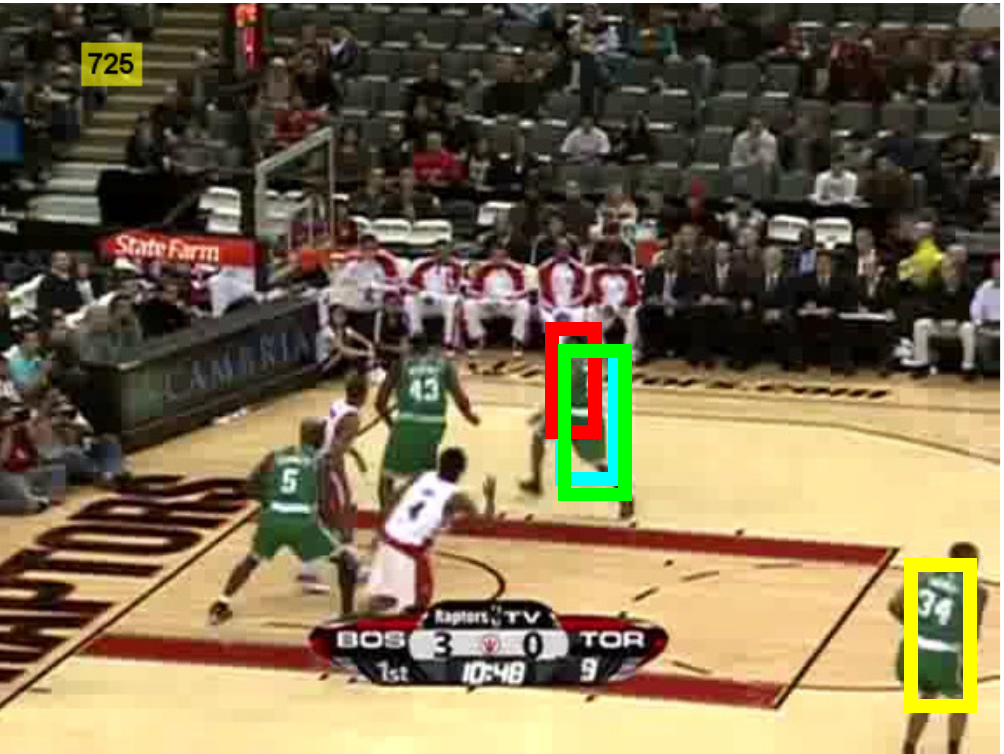}}  \vspace{4.pt}
\end{minipage}

  \begin{minipage}[b]{0.2\linewidth}
  \centering
  \centerline{\includegraphics[width=\linewidth]{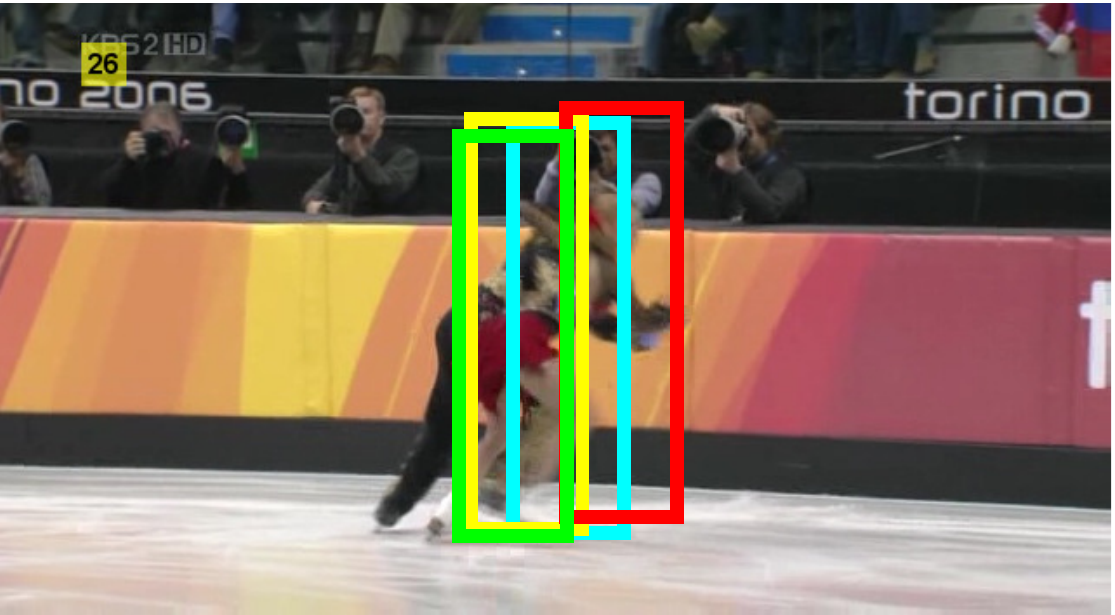}}\vspace{4.pt}
\end{minipage}
\begin{minipage}[b]{0.2\linewidth}
  \centering
  \centerline{\includegraphics[width=\linewidth]{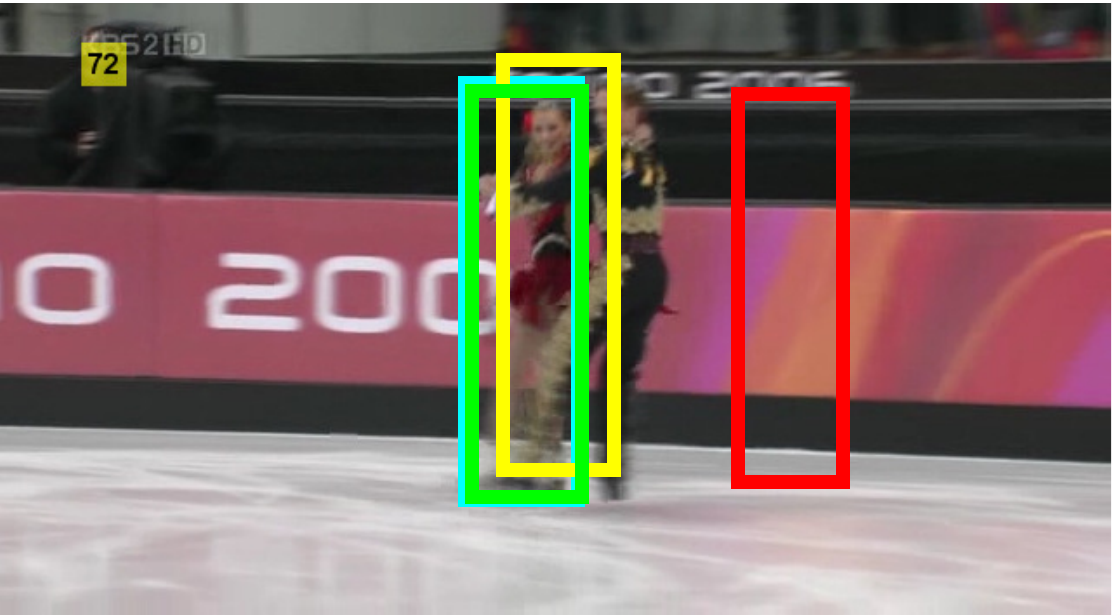}}\vspace{4.pt}
\end{minipage}
\begin{minipage}[b]{0.2\linewidth}
  \centering
  \centerline{\includegraphics[width=\linewidth]{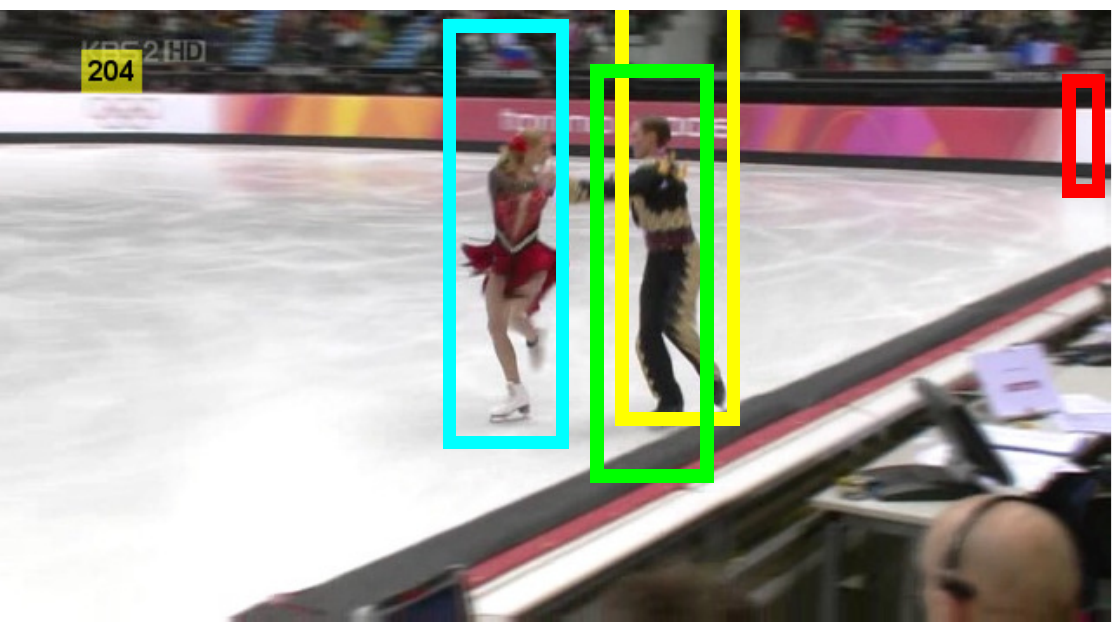}}\vspace{4.pt}
\end{minipage}
\begin{minipage}[b]{0.2\linewidth}
  \centering
  \centerline{\includegraphics[width=\linewidth]{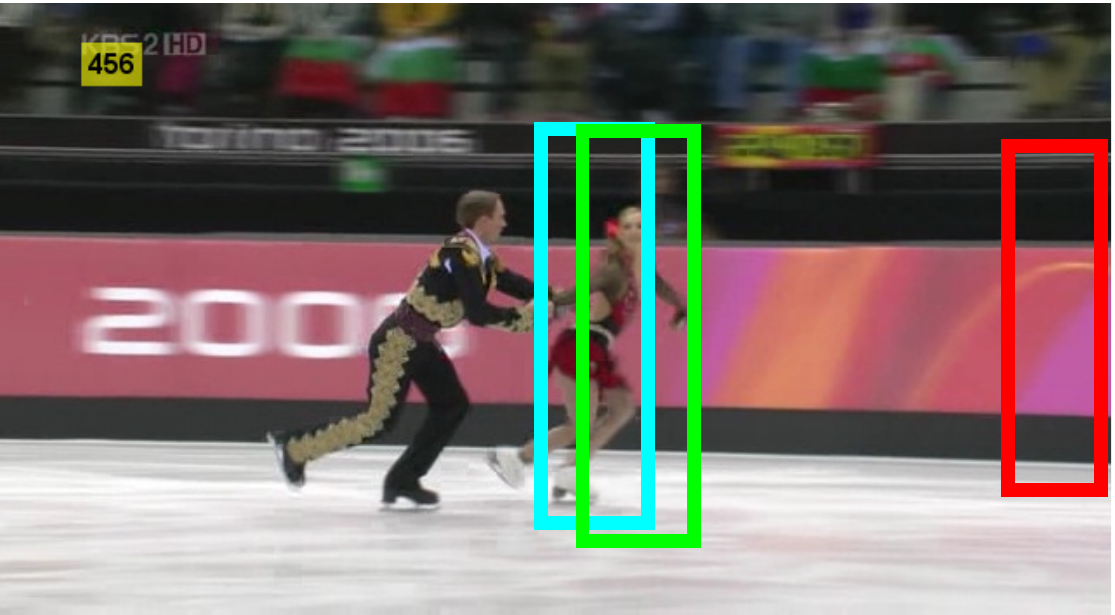}}  \vspace{4.pt}
\end{minipage}

  \begin{minipage}[b]{0.2\linewidth}
  \centering
  \centerline{\includegraphics[width=\linewidth]{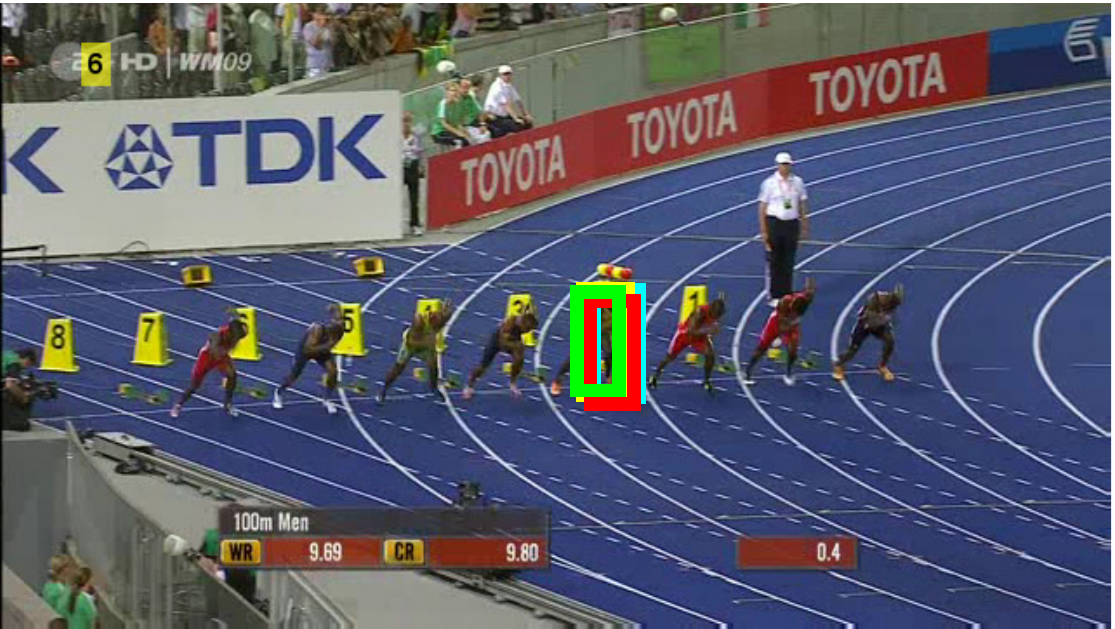}}
\end{minipage}
\begin{minipage}[b]{0.2\linewidth}
  \centering
  \centerline{\includegraphics[width=\linewidth]{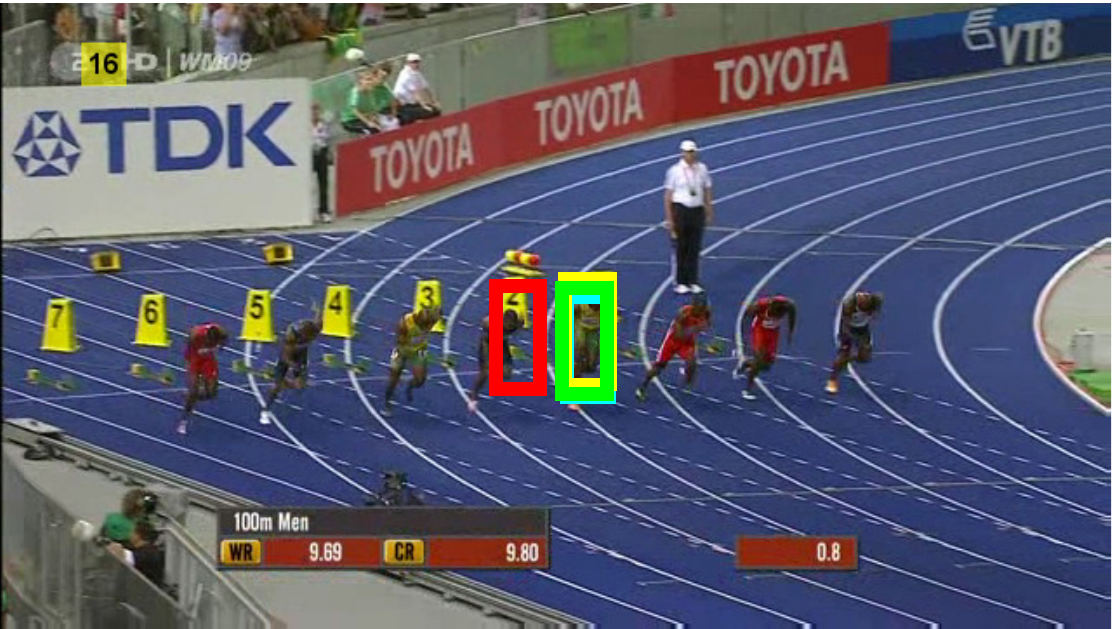}}
\end{minipage}
\begin{minipage}[b]{0.2\linewidth}
  \centering
  \centerline{\includegraphics[width=\linewidth]{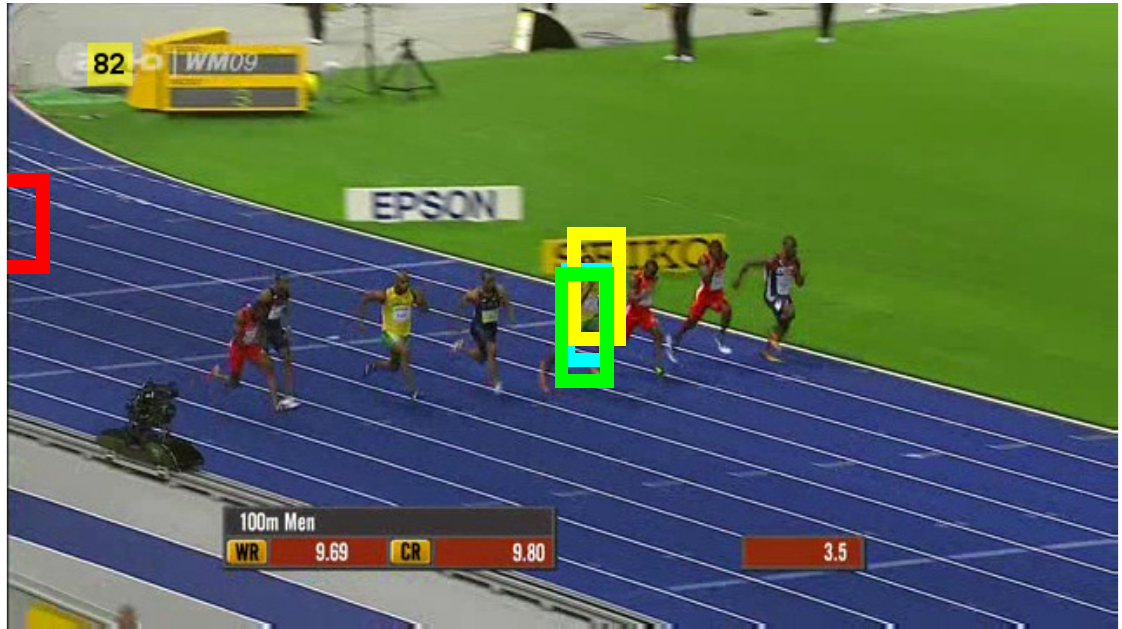}}
\end{minipage}
\begin{minipage}[b]{0.2\linewidth}
  \centering
  \centerline{\includegraphics[width=\linewidth]{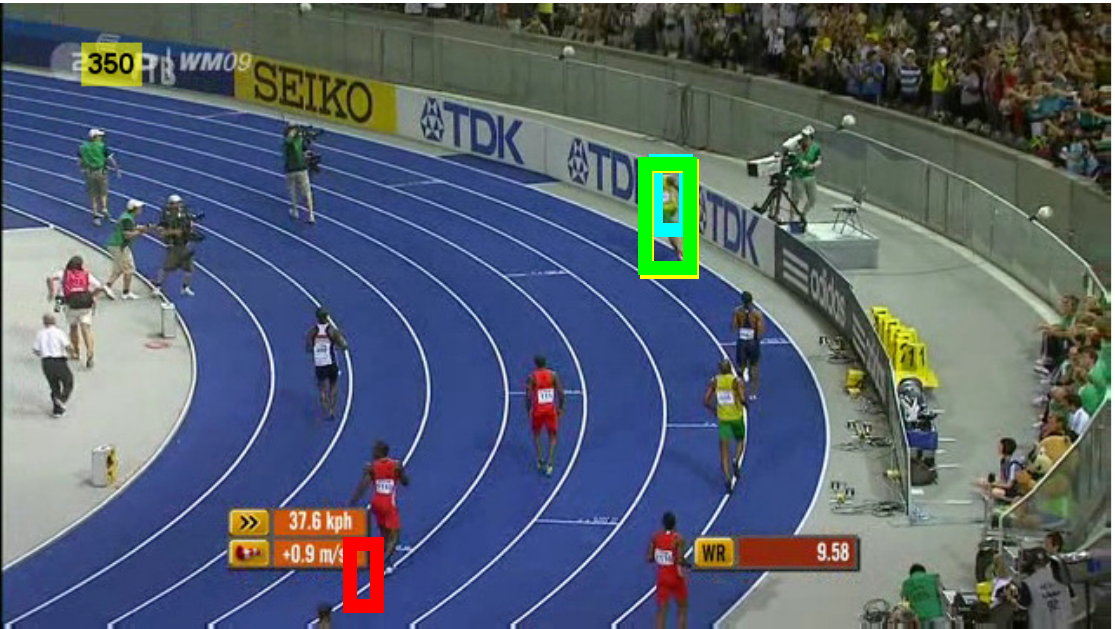}}  
\end{minipage}
\caption{Screen shots obtained by four top ranked trackers in the tracking of Bird1, Basketball, Skating2, and Bolt videos. Cyan box: Ours; Yellow box: robust superpixels tracking (SPT); Red box: discriminative tracking using tensor pooling (TPT); Green box: correlation-filter based scale-adaptive visual tracking with hybrid-scheme sample learning (KCFS).}
\label{fig4}
\end{figure*}

\subsection{Incremental Positive and Negative Subspaces Learning}
The update and learning scheme used in SPTPT is based on the incremental subspace learning \cite{IRTSA1}. In order to make the tracker more robust against drifts, we refer to papers \cite{TPT2, TPT} to introduce discriminative framework called negative subspace learning into the learning scheme. Hence, the learning scheme used in SPTPT is called incremental positive and negative subspaces learning. 

If the algorithm reaches the update rate $u$ (in this paper, $u$ is set to 5), then a 3-order tensor $\mathcal T \in \textbf{R}^{z \times s \times u}$ corresponding to positive subspace is constructed. We use the incremental rank tensor subspace analysis (IRTSA) algorithm \cite{IRTSA1} to find the dominant projection subspaces of it. The details of IRTSA can be referred in \cite{IRTSA1}. To learn the positive subspace, it is necessary to evaluate the likelihood between the candidate sample and its approximation in the learned positive subspace. Given a third order tensor $\mathcal J \in \textbf{R}^{z \times s \times 1}$ of a candidate in the new frame, the evaluation of its likelihood in the learned positive subspace can be determined by the reconstruction error as follows,
\begin{equation}
RE_1 = \sum_{i=1}^{2}||(\emph{\textbf{J}}_{(i)} - \emph{\textbf{M}}_{(i)})-(\emph{\textbf{J}}_{(i)} - \emph{\textbf{M}}_{(i)})\prod_{j=1}^{2} \times_{j} (U^{(j)} \cdot U^{(j)\mathrm T})||^2,
\label{fun1}
\end{equation}
\begin{equation}
RE_2 = ||(\emph{\textbf{J}}_{(3)} - \emph{\textbf{M}}_{(3)}) -  (\emph{\textbf{J}}_{(3)} - \emph{\textbf{M}}_{(3)}) \cdot (V^{(3)} \cdot V^{(3)\mathrm T}) ||^2,
\label{fun2}
\end{equation}
\begin{equation}
RE = \gamma RE_1 + (1 - \gamma) RE_2,
\label{fun3}
\end{equation}
where, $\mathcal M$ is the mean tensor of $\mathcal T$, $\emph{\textbf{M}}_{(i)} (i = 1,2,3)$ is the mode-$i$ unfolding matrix of $\mathcal M$, $\emph{\textbf{J}}_{(i)} (i = 1,2,3)$ is the mode-$i$ unfolding matrix of $\mathcal J$, and $\gamma$ is the control weight, in this paper, $\gamma = 0.5$.

As to the negative subspace learning, in contrast to the positive subspace learning which collects the positive samples per frame using the tensor-pooled sparse features obtained from the tracked frames and is incremental learned by IRTSA, the negative samples are only collected in the last tracked frame through extracting superpixels a certain distance threshold (several pixels) around the estimated location of the target \cite{TPT2, TPT}. Since these negative samples are collected in only one frame rather than a sequence of frames, the negative subspace is learned directly by doing tensor decomposition (TD) of the sparse pooling tensors of these samples. The likelihood can also be calculated by Equations (\ref{fun1})-(\ref{fun3}).

\begin{figure}
\centering 
\centerline{\includegraphics[width=0.8\columnwidth]{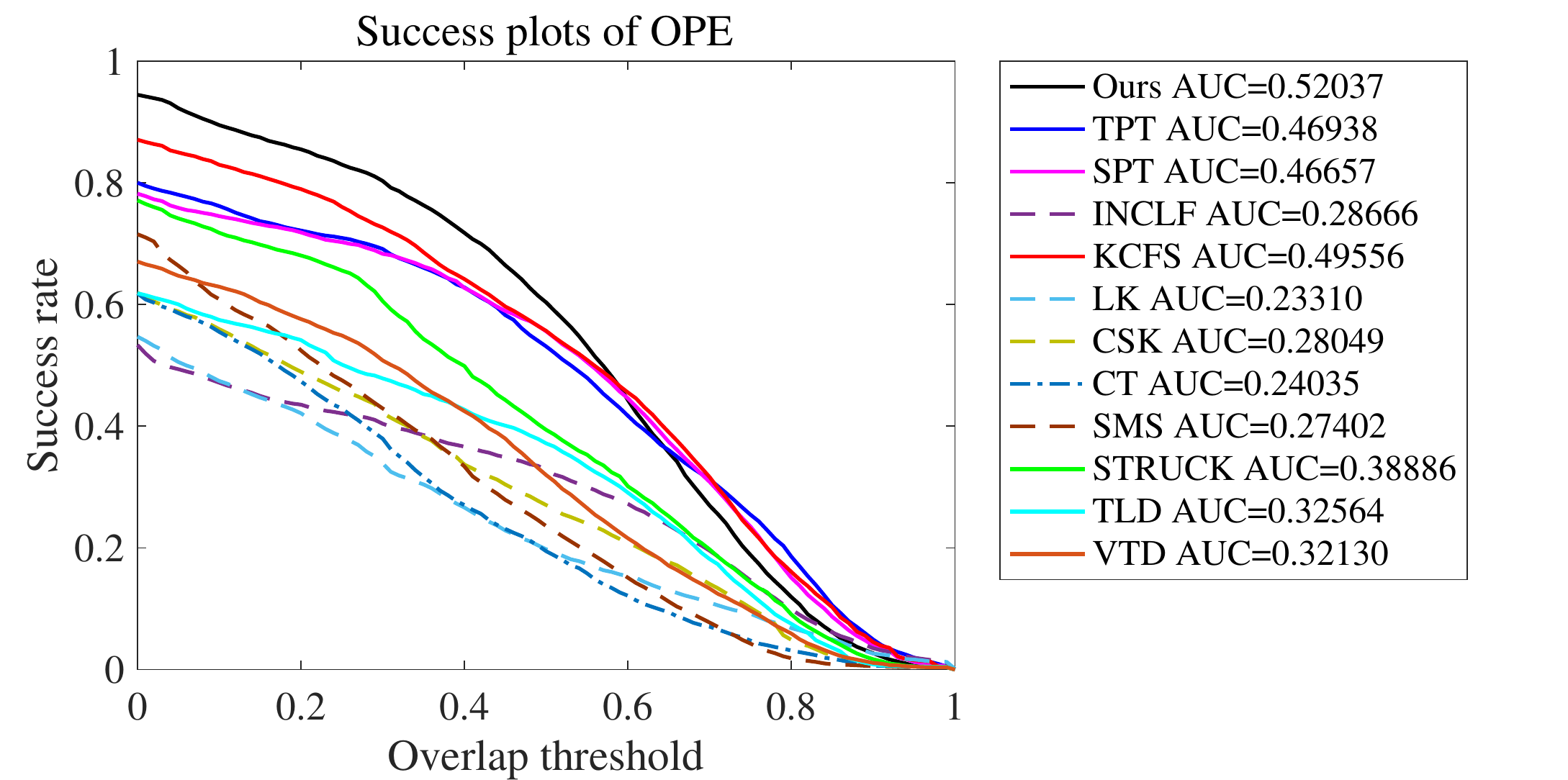}}
\caption{Overall success plots of 12 trackers. Rank: Ours, KCFS, TPT, SPT, STRUCK, TLD, VTD, INCLF, CSK, SMS, CT, LK.}
\label{fig2}
\end{figure}

\begin{figure}
\centering 
\centerline{\includegraphics[width=0.8\columnwidth]{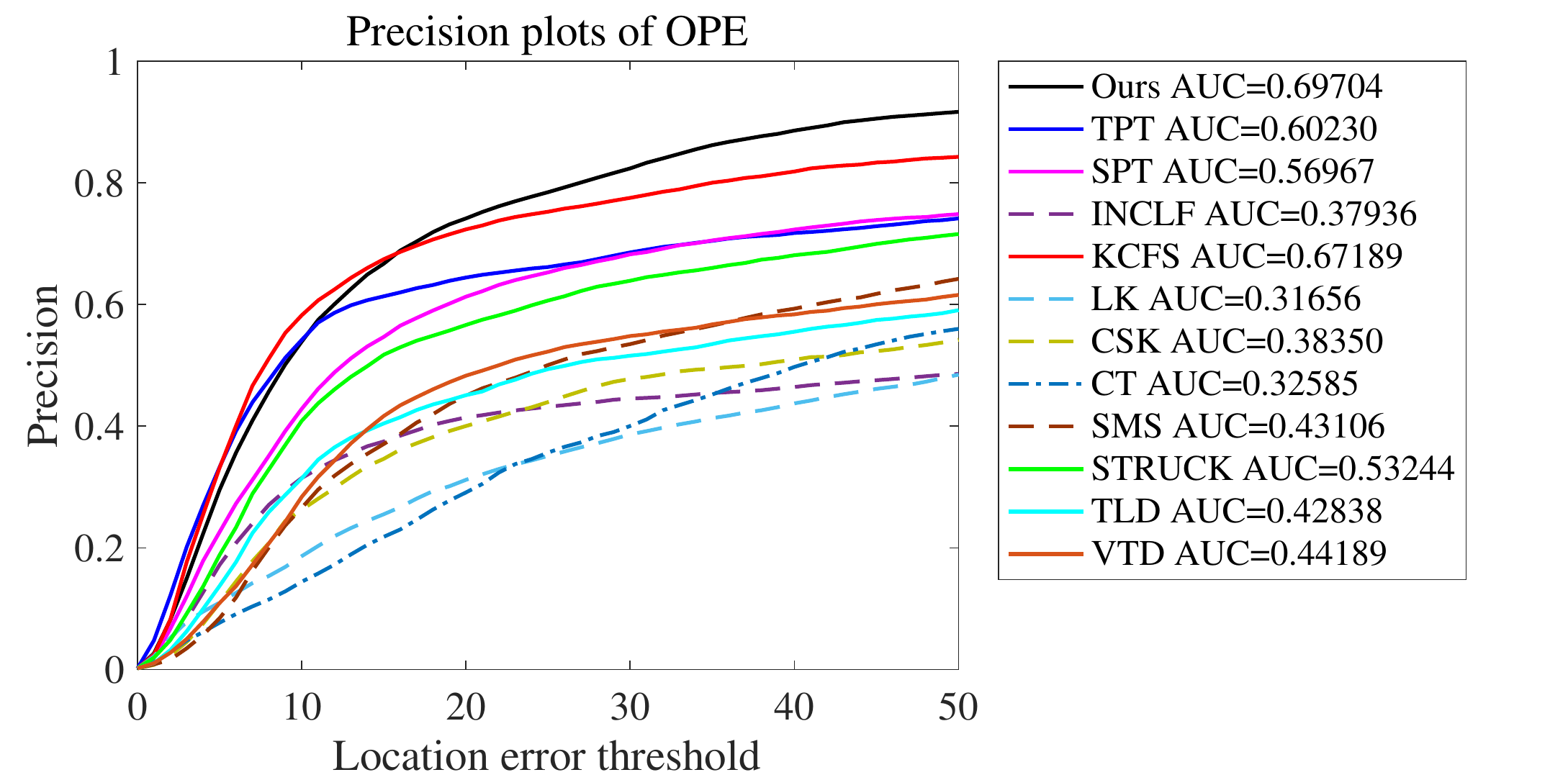}}
\caption{Overall precision plots of 12 trackers. Rank: Ours, KCFS, TPT, SPT, STRUCK, VTD, SMS, TLD, CSK, INCLF, CT, LK.}
\label{fig3}
\end{figure}

\subsection{Motion Model based on Bayesian Inference}
The motion model of SPTPT is based on Bayesian inference. Let $X_t = \{x_t, y_t, \vartheta_t, s_t, \beta_t, \phi_t\}$ represent the state (affine transformation parameters) at $t$th frame, where $x_t$ is the \emph{x} translation, $y_t$ is the \emph{y} translation, $\vartheta_t$ is the rotation angle, $s_t$ is the scale, $\beta_t$ is the aspect ratio, and $\phi_t$ is the skew direction, let $ \mathcal S_t$ represent a set of the observations $\{S_1, S_2, ...,S_t\}$ at time $t$. The posterior probability is calculated as follow,
\begin{equation}
p(X_t|\mathcal S_t) \propto p(S_t|X_t) \int p(X_t|X_{t-1})p(X_{t-1}|\mathcal{S}_{t-1})dX_{t-1},
\label{fun05}
\end{equation}
where, $p(S_t | X_t)$ represents the observation model here is the likelihood function, and $p(X_t|X_{t-1})$ denotes the dynamic model between states $X_t$ and $X_{t-1}$. We apply a particle filter \cite{PF} to generate samples (number of positive samples: 600 and number of negative samples: 200, in this paper) through estimating the distribution. The optimal state can be obtained by using the maximum a posteriori (MAP) estimation,
\begin{equation}
\hat{X}_t = \arg \max\limits_{X_t^i} p(S_t^i | X_t^i)p(X_t^i|X_{t-1}), i = 1,2,...,b,
\label{fun06}
\end{equation}
where, $b$ is the number of samples and $X_t^i$ represents the sample $i$ of state $X_t$.
The dynamic model $p(X_t|X_{t-1})$ is formulated using the random walk as follow \cite{TPT},
\begin{equation}
p(X_t|X_{t-1}) = \mathcal N (X_t; X_{t-1},\Omega),
\label{fun07}
\end{equation}
where, $\Omega$ is a diagonal covariance matrix. Its diagonal elements are $ \sigma_{x}^2, \sigma_y^2, \sigma_\vartheta^2, \sigma_s^2, \sigma_\beta^2, \sigma_\phi^2$, respectively. 

Finally, the likelihood of a candidate in both positive and negative subspaces is formulated as follow: 
\begin{equation}
p(Y_t|X_t) \propto  \exp(RE^{(-)} - RE^{(+)}).
\label{fun08}
\end{equation}

To make the tracker more robust and avoid overfitting, SPTPT uses the likelihood function above to control the learning: only when the best candidate's likelihood $exp(RE^{(-)} - RE^{(+)}) > \Phi$, it can be accepted into the updating scheme, where $\Phi$ is a threshold, in this paper, it is set to 0. 

\subsection{Algorithm Summary}
Fig. \ref{fig1} illustrates the workflow of the proposed tracking algorithm:
\begin{itemize}
  \item [1)] 
  Use particle filter (PF) and affine transformation to produce some candidates.
  \item [2)]
  Use simple non-iterative clustering (SNIC) to generate superpixels.
  \item [3)]
  Compute the histograms of color and spatial information in each superpixel.
  \item [4)]
  Construct the histograms to a third order tensor.
  \item [5)]
  Continue \textbf{Steps 3-4} until the features of all superpixels of all candidates are obtained.
  \item [6)]
  Determine whether the current frame is the first frame. If yes, produce the dictionary matrix $\emph{\textbf{D}}$, store the tensor $\mathcal{T}$ of it into an updating sequence and then go to \textbf{Step 1}, otherwise go to \textbf{Step 7}.
  \item [7)]
  Use $\emph{\textbf{D}}$ to do the pooling of tensors obtained in \textbf{Step 3} to obtain the sparse pooling tensors $\mathcal J$.
  \item [8)]
  Evaluate the likelihood of $\mathcal J$ in positive and negative subspaces using Equation (\ref{fun08}).
  \item [9)]
  According to the likelihood, update the discriminative appearance model. If the max likelihood $>$ 0, store the tensor $\mathcal J$ of max likelihood into the updating sequence and use PF to draw some negative samples to update negative subspace and if algorithm reaches update rate $u$, use IRTSA to update positive subspace.
  \end{itemize}
Continue \textbf{Steps 1-9} until the last frame is processed.

\section{Experiments and results}
\label{sec:eres}
To validate the proposed method, we follow the protocol of the benchmark \cite{Bench}. We compared our method with eleven state-of-the-art methods: discriminative tracking using tensor pooling (TPT) \cite{TPT}, robust superpixel tracking (SPT) \cite{SPT1}, correlation-filter based scale-adaptive visual tracking with hybrid-scheme sample learning (KCFS) \cite{KCFS}, inverse nonnegative local coordinate factorization (INCLF) \cite{INCLF}, optical flow method using observation model based on intensity restrictions (LK) \cite{LK}, structured output tracking with kernels (STRUCK) \cite{STRUCK}, P-N learning: bootstrapping binary classifiers by structural constraints (TLD) \cite{TLD}, visual tracking decomposition (VTD) \cite{VTD}, circulant structure of tracking-by-detection with kernels (CSK) \cite{CSK}, mean-shift blob tracking through scale space (SMS) \cite{SMS}, real-time compressive tracking (CT) \cite{CT} and tested them on 24 sequences (Basketball, Biker, Bird1, Bird2, Bolt, CarScale, DragonBaby, Football1, Lemming, Liquor, MountainBike, Panda, RedTeam, Rubik, Singer1, Skating1, Skating2, Soccer, Subway, Surfer, Tiger2, Trans, Trellis, Woman) with multiple visual tracking challenges (illumination variation, scale variation, occlusion, deformation, motion blur, fast motion, in-plane rotation, out-of-plane rotation, out-of-view, background clutters, and low resolution) from the benchmark \cite{Bench}. SPTPT is implemented using MATLAB and conducted on a Mac with OSX 10.14, Intel Core i5 2.3 GHz 4 cores CPU, and 16 GB RAM. To keep the fairness, all methods are used one default parameter for all sequences without any tuning, that is to say, the results shown in this section are the lower-bound performance of each method.

\subsection{Evaluation Metrics}
Precision plots and success plots \cite{Bench} are used to evaluate the overall performance and robustness of trackers. A precision plot illustrates the ratio of frames whose center location error is within a threshold distance to the ground truth. A success plot illustrates the percentage of frames of which overlapping rate between tracked results and ground truth is larger than a certain threshold. The final rank is according to the area under the curve (AUC) of each tracker. The precision and success rate of all trackers are tested under one-pass evaluation (OPE).
\begin{figure*}[h]
\centering
\begin{minipage}[b]{0.3\linewidth}
  \centering
  \centerline{\includegraphics[width=\linewidth]{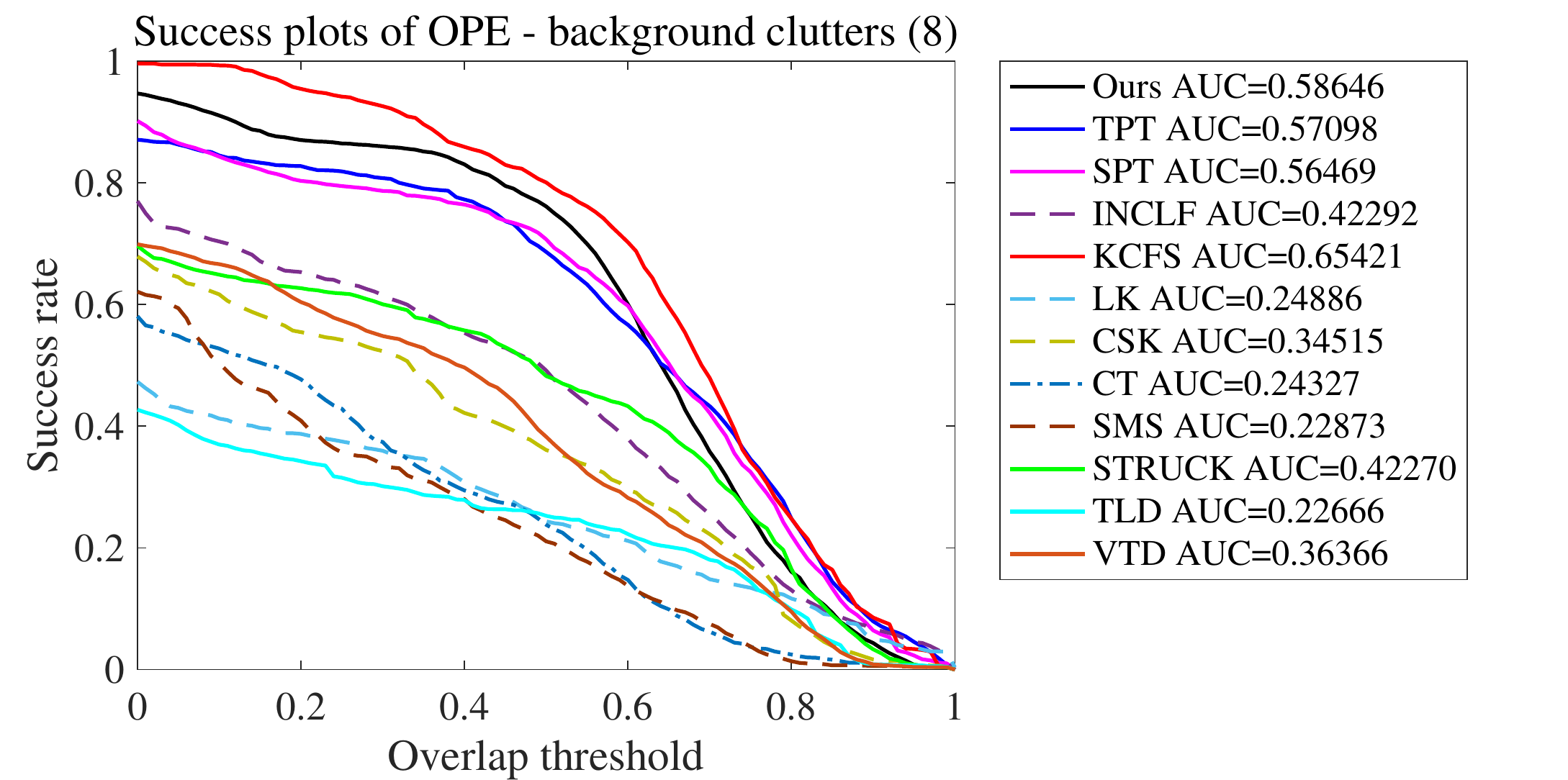}}\vspace{4.pt}
\end{minipage}
\begin{minipage}[b]{0.3\linewidth}
  \centering
  \centerline{\includegraphics[width=\linewidth]{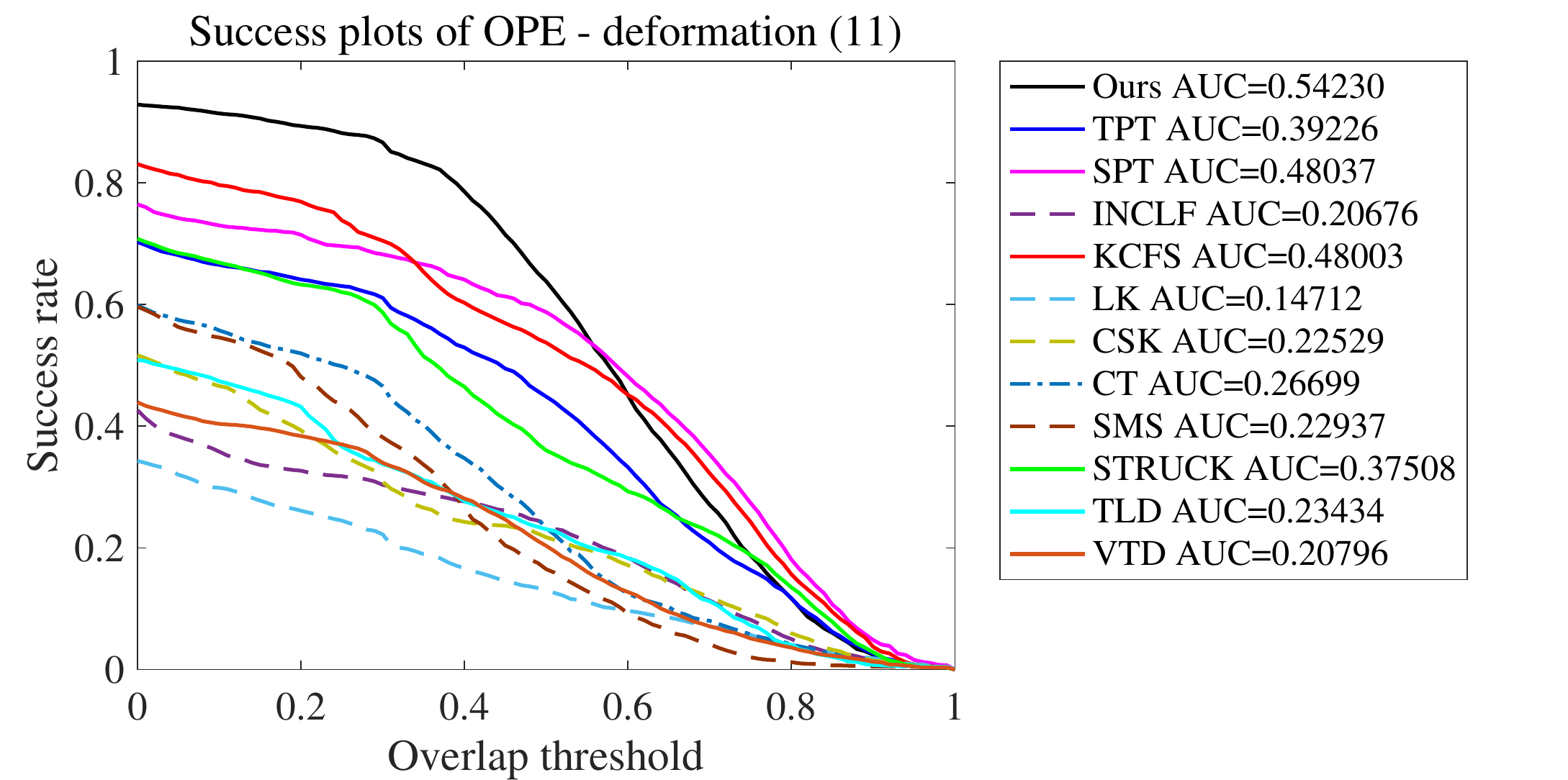}}\vspace{4.pt}
\end{minipage}
\begin{minipage}[b]{0.3\linewidth}
  \centering
  \centerline{\includegraphics[width=\linewidth]{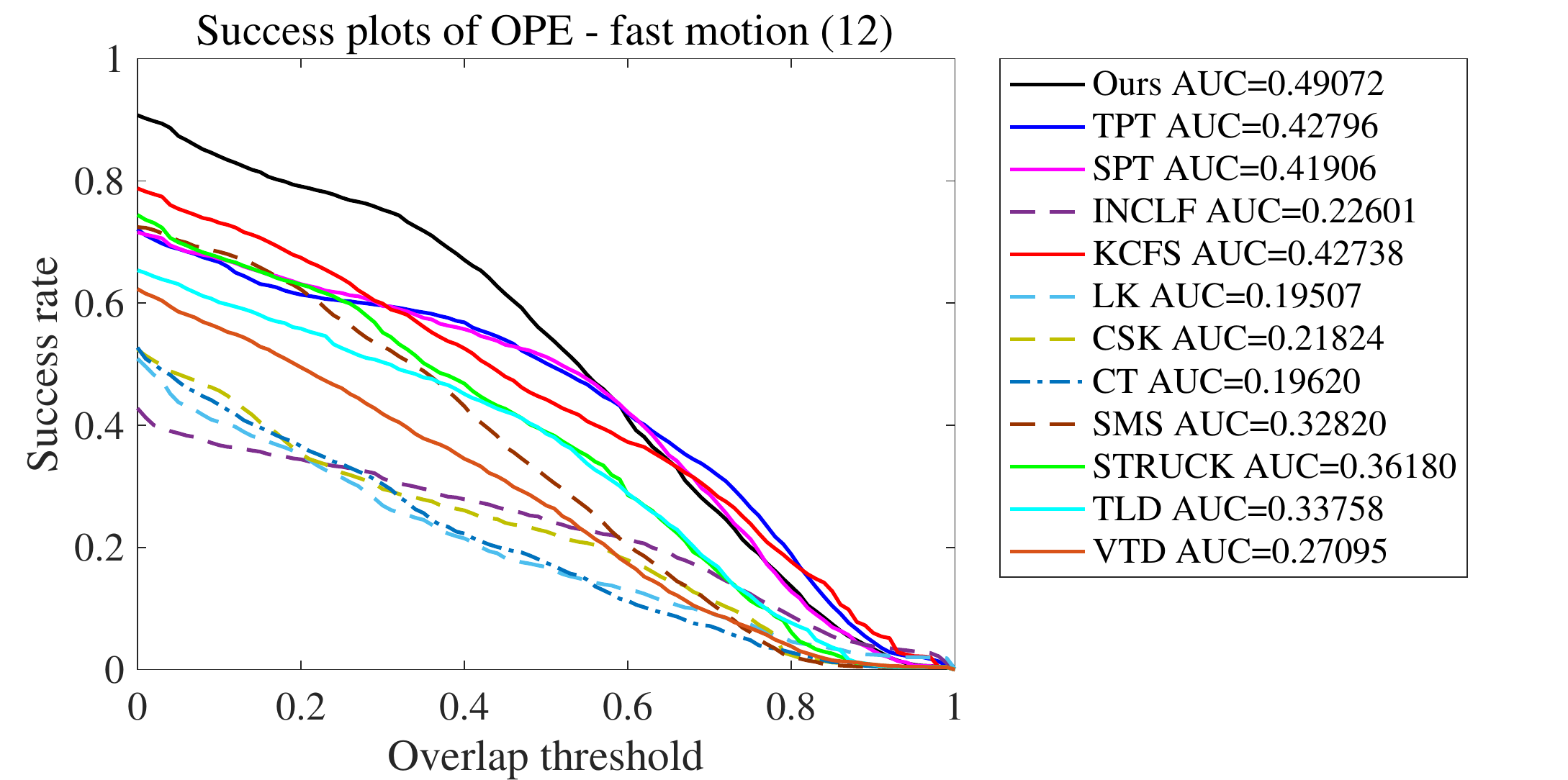}}\vspace{4.pt}
\end{minipage}
\begin{minipage}[b]{0.3\linewidth}
  \centering
  \centerline{\includegraphics[width=\linewidth]{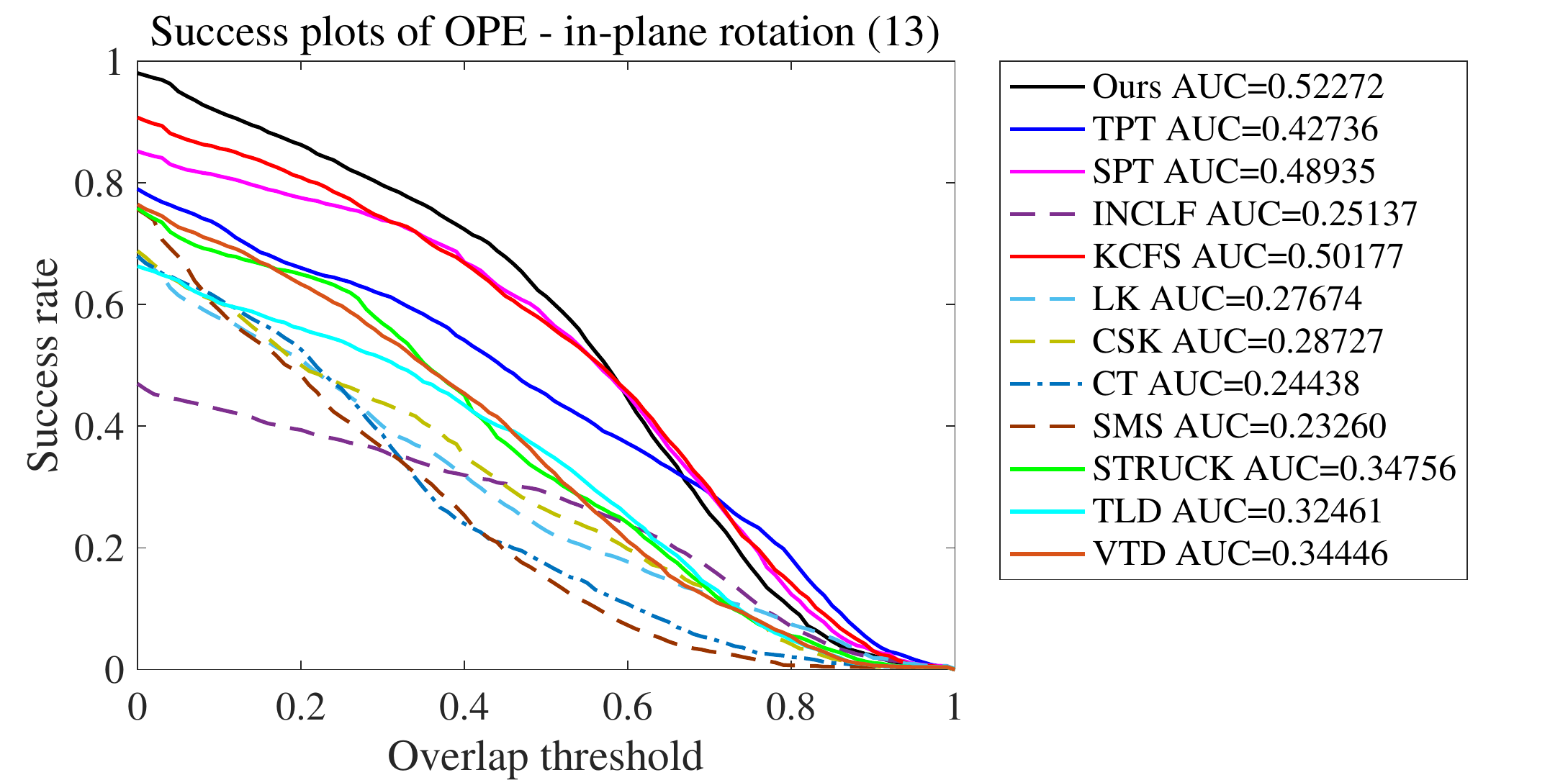}}  \vspace{4.pt}
  \end{minipage}
  \begin{minipage}[b]{0.3\linewidth}
  \centering
  \centerline{\includegraphics[width=\linewidth]{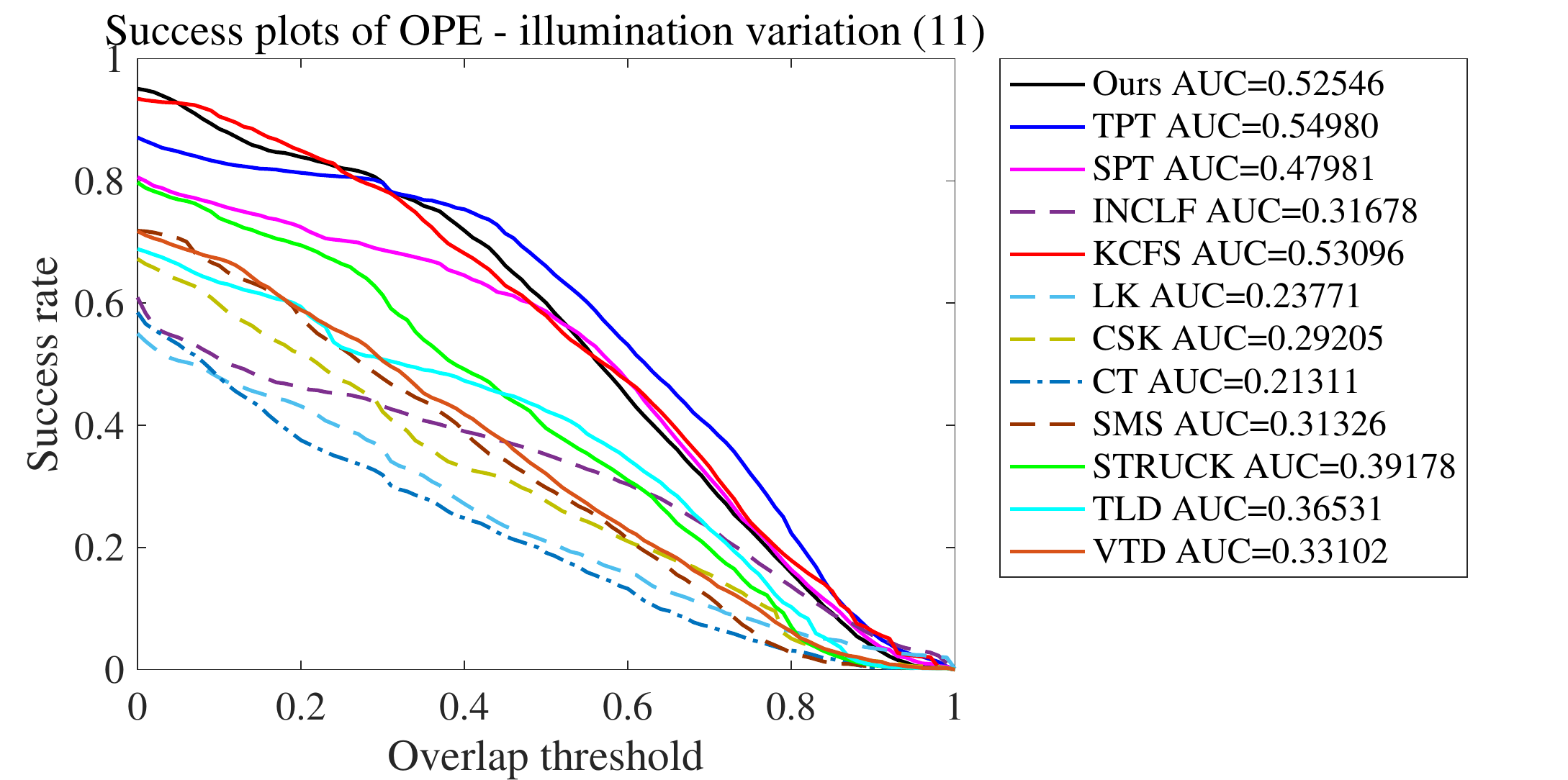}}\vspace{4.pt}
\end{minipage}
\begin{minipage}[b]{0.3\linewidth}
  \centering
  \centerline{\includegraphics[width=\linewidth]{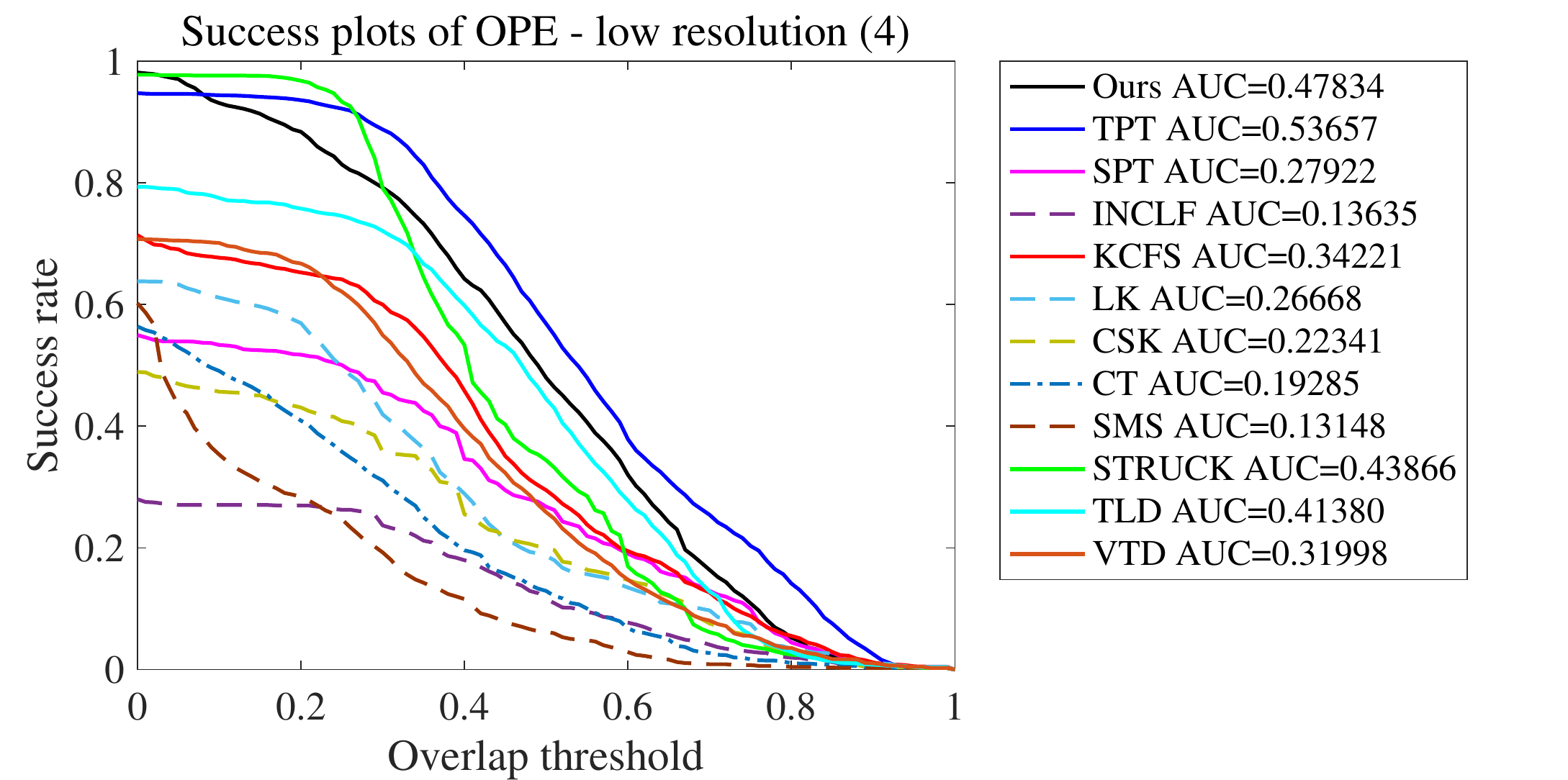}}\vspace{4.pt}
\end{minipage}
\begin{minipage}[b]{0.3\linewidth}
  \centering
  \centerline{\includegraphics[width=\linewidth]{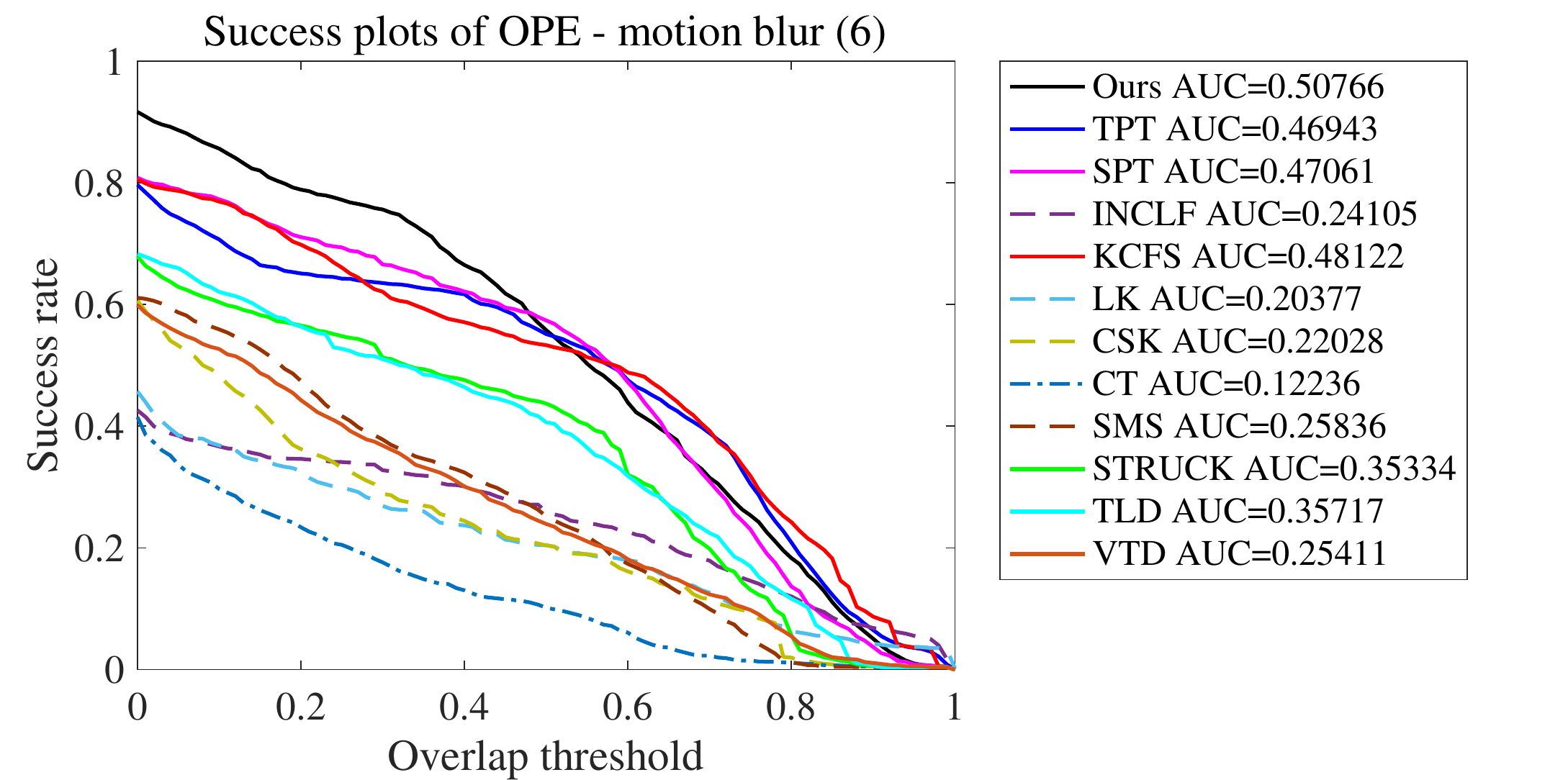}}\vspace{4.pt}
\end{minipage}
\begin{minipage}[b]{0.3\linewidth}
  \centering
  \centerline{\includegraphics[width=\linewidth]{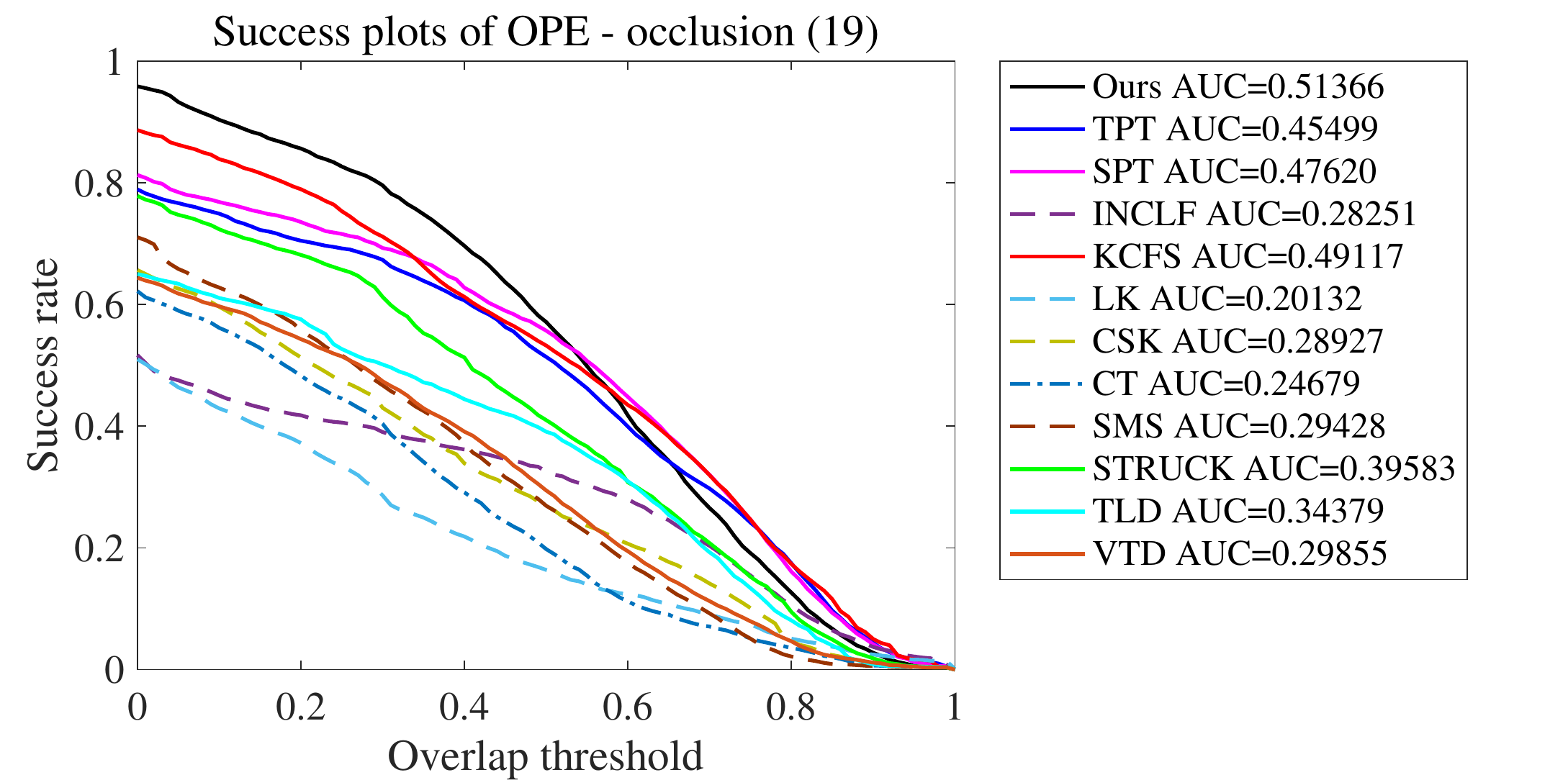}}  \vspace{4.pt}
\end{minipage}
  \begin{minipage}[b]{0.3\linewidth}
  \centering
  \centerline{\includegraphics[width=\linewidth]{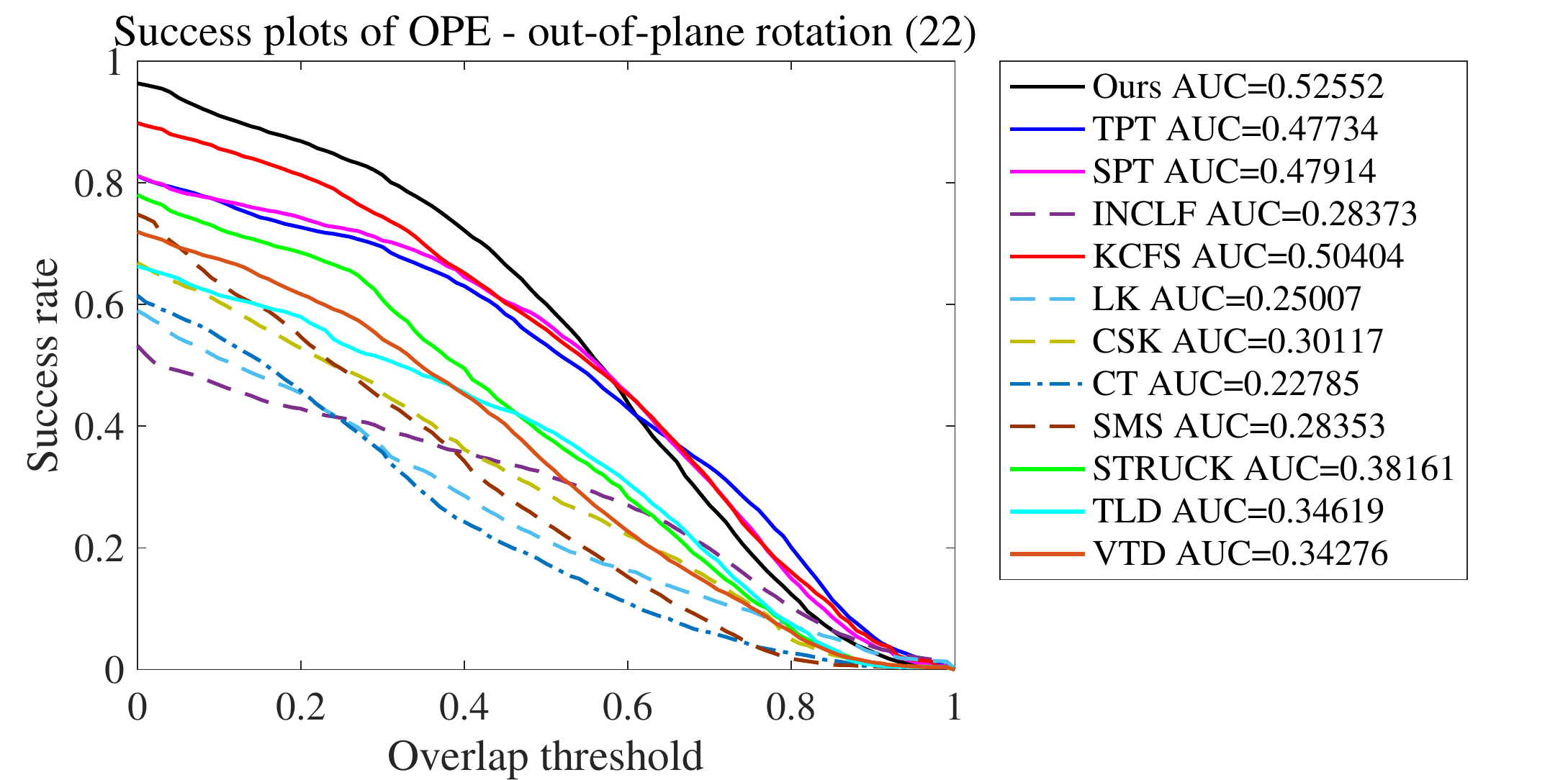}}\vspace{4.pt}
\end{minipage}
\begin{minipage}[b]{0.3\linewidth}
  \centering
  \centerline{\includegraphics[width=\linewidth]{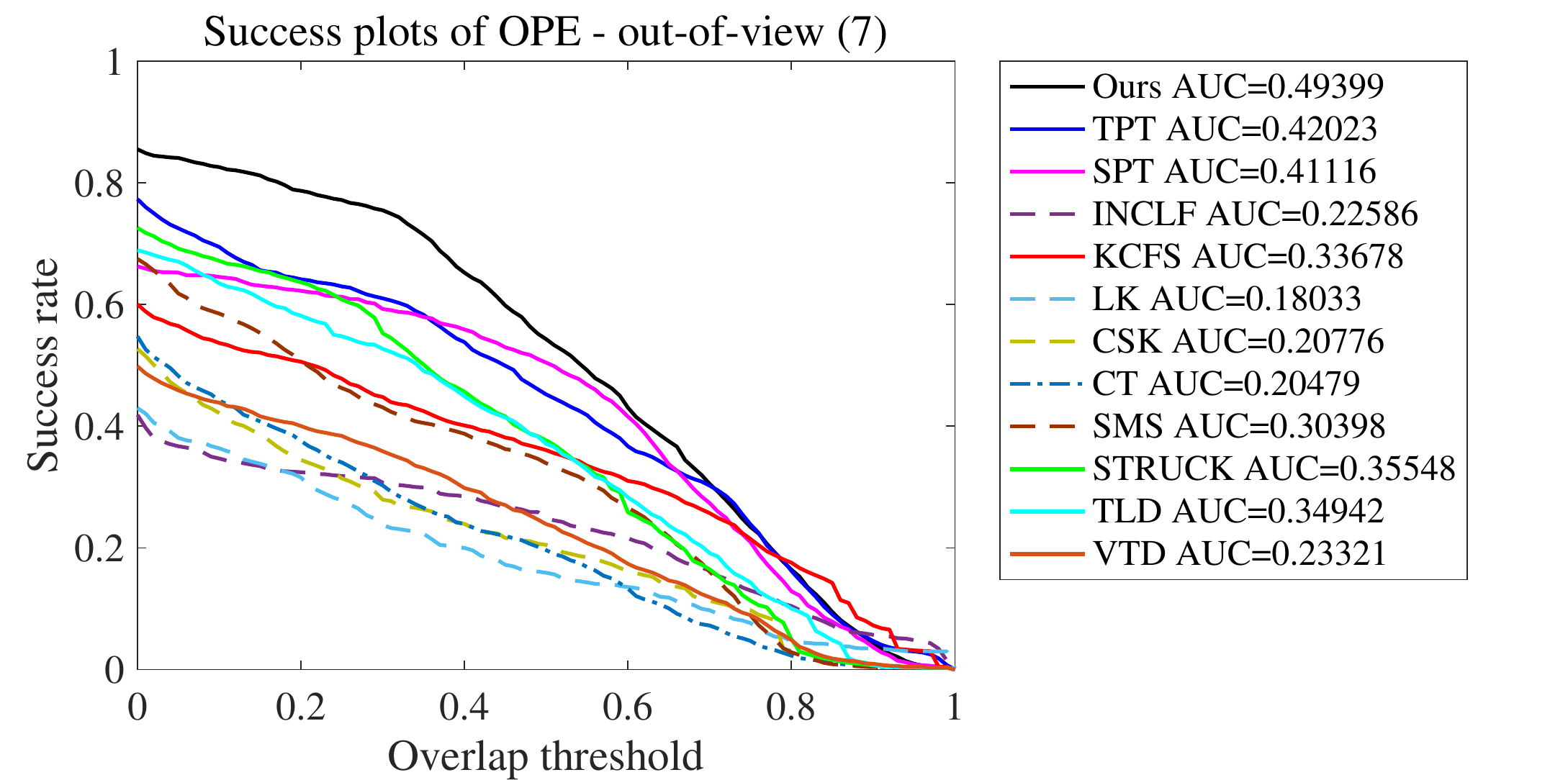}}\vspace{4.pt}
\end{minipage}
\begin{minipage}[b]{0.3\linewidth}
  \centering
  \centerline{\includegraphics[width=\linewidth]{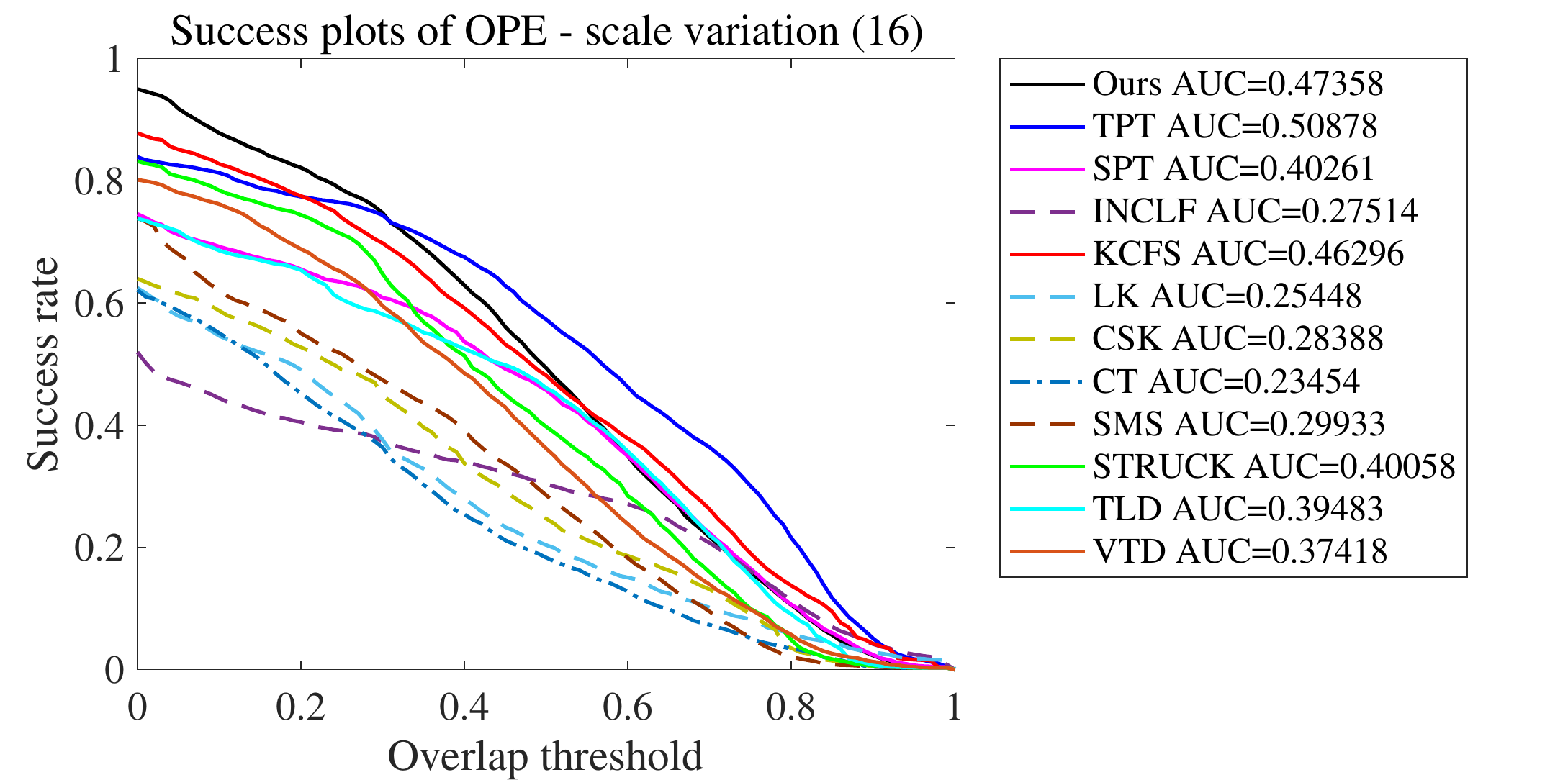}}\vspace{4.pt}
\end{minipage}
\caption{The success plots of 12 trackers about 11 challenging attributes on the benchmark.}
\label{fig6}
\end{figure*}

\subsection{Performance Analysis}
\begin{figure*}[h]
\centering
\begin{minipage}[b]{0.3\linewidth}
  \centering
  \centerline{\includegraphics[width=\linewidth]{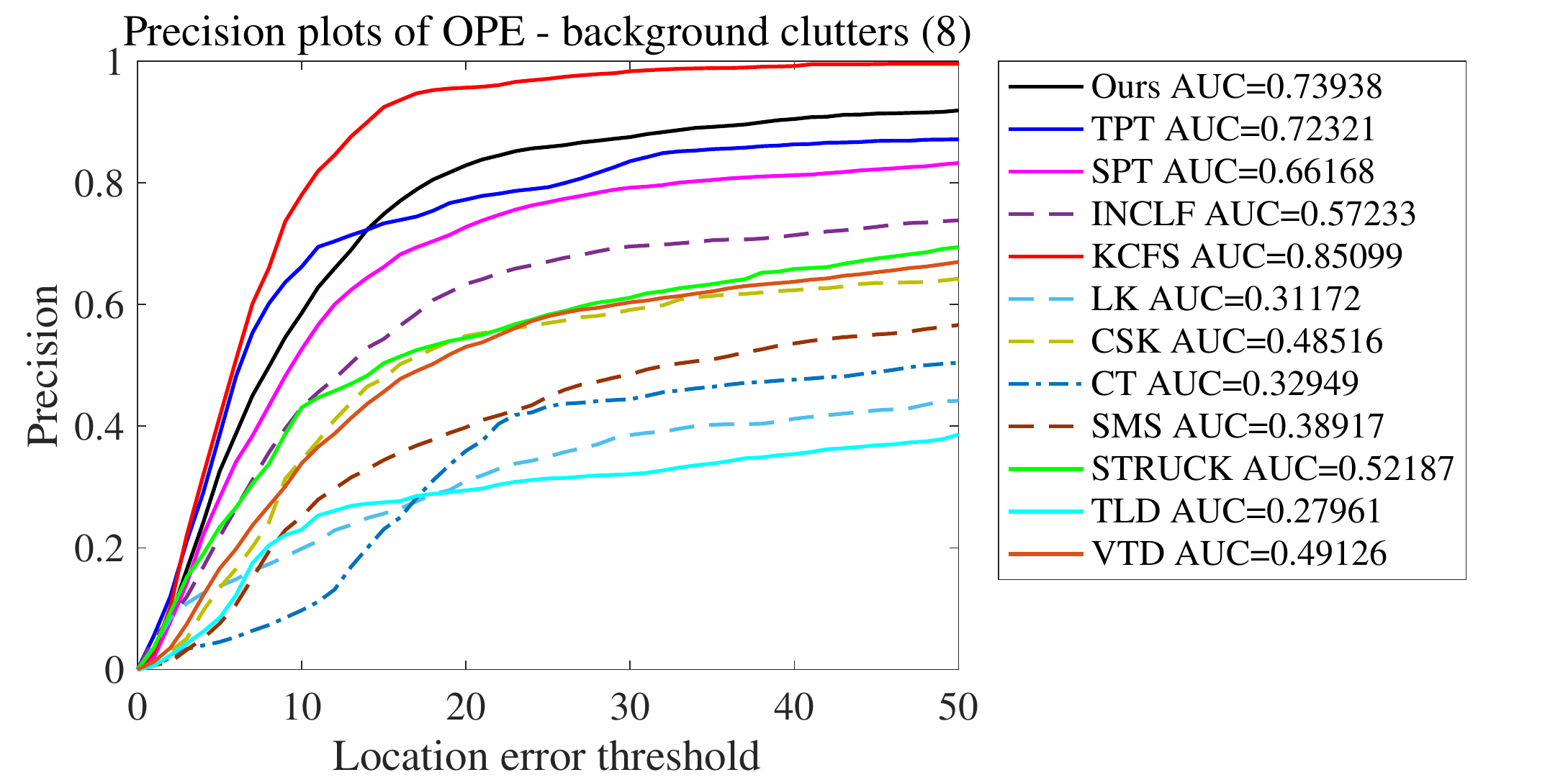}}\vspace{4.pt}
\end{minipage}
\begin{minipage}[b]{0.3\linewidth}
  \centering
  \centerline{\includegraphics[width=\linewidth]{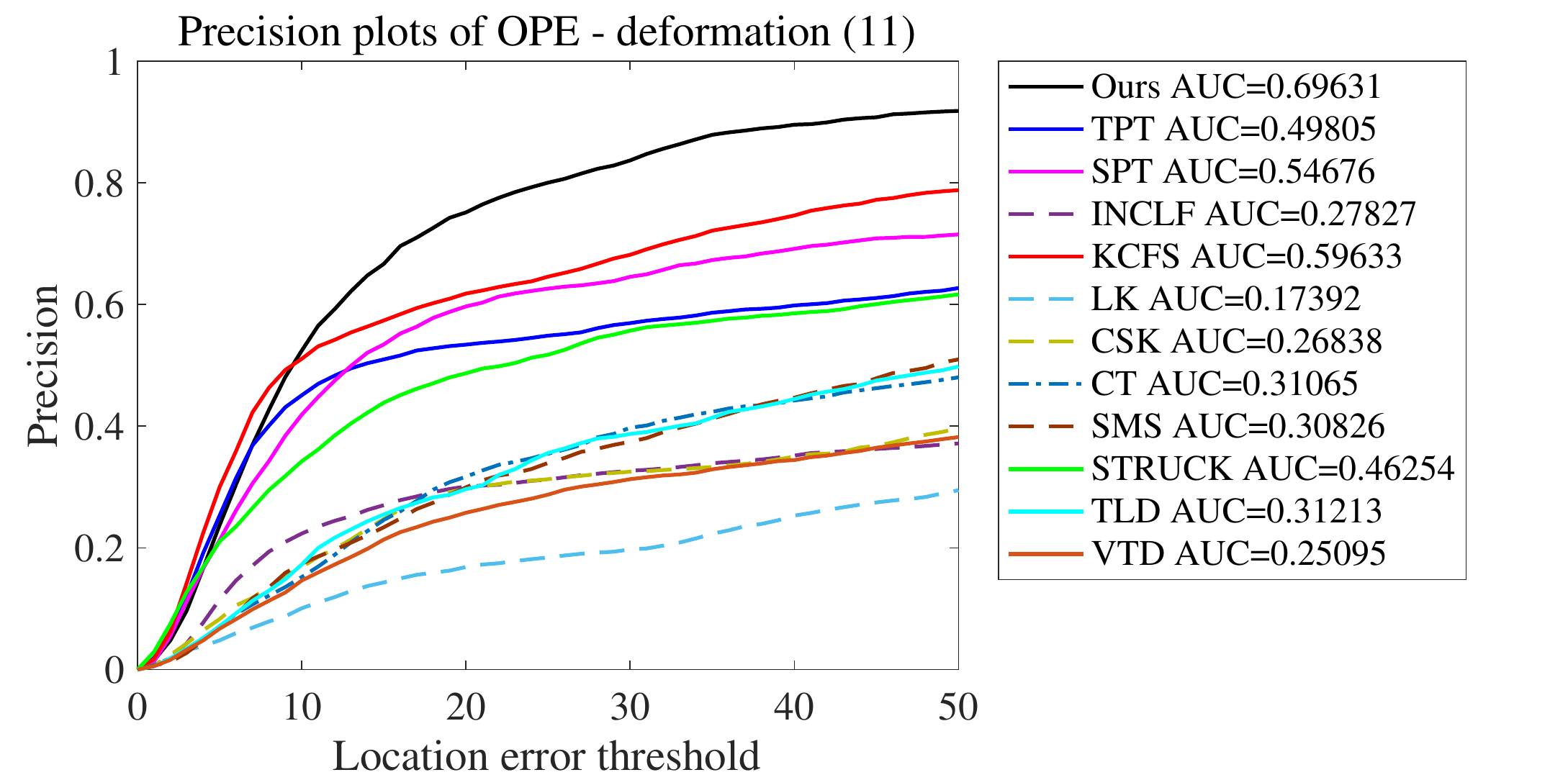}}\vspace{4.pt}
\end{minipage}
\begin{minipage}[b]{0.3\linewidth}
  \centering
  \centerline{\includegraphics[width=\linewidth]{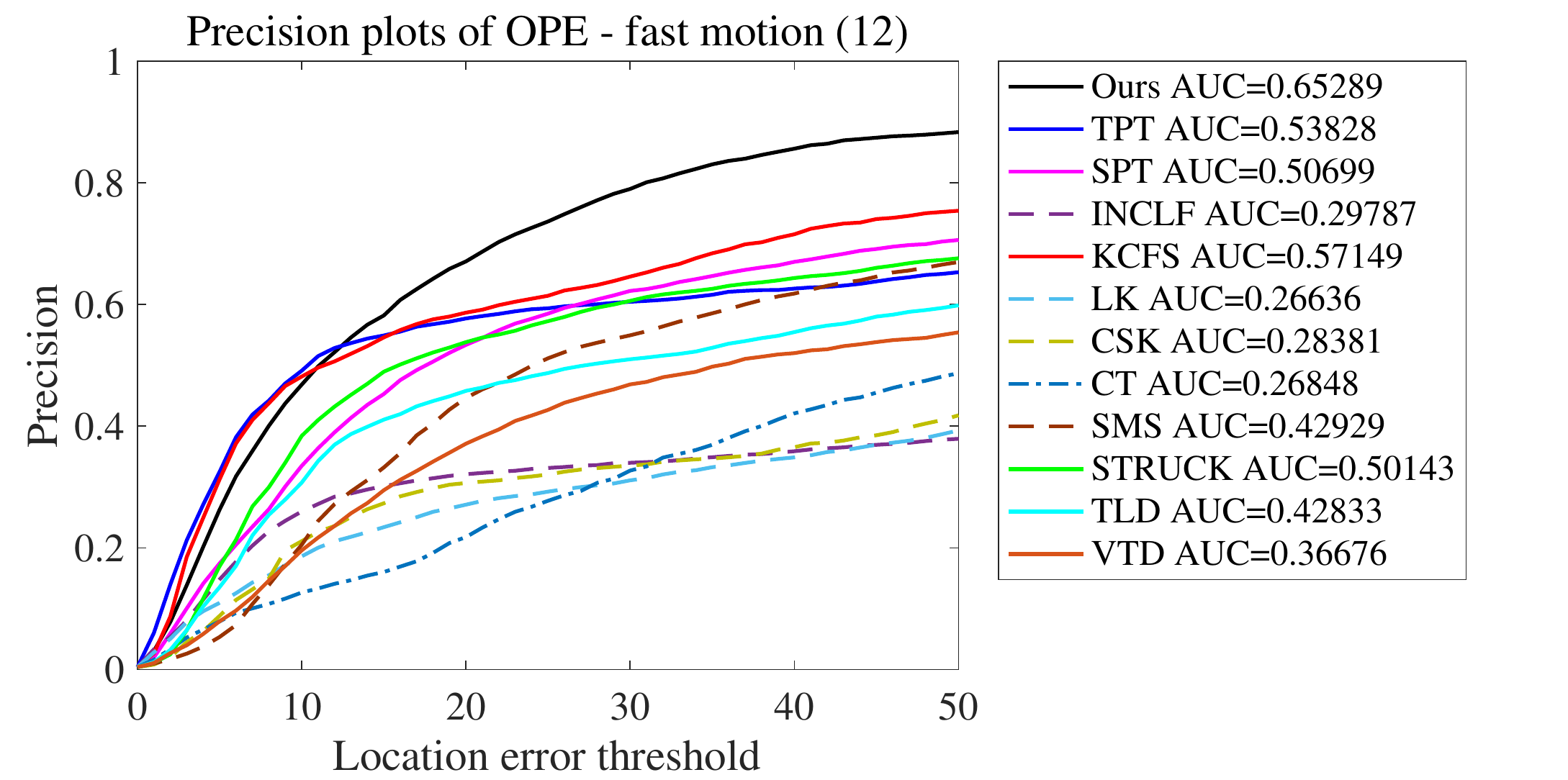}}\vspace{4.pt}
\end{minipage}
\begin{minipage}[b]{0.3\linewidth}
  \centering
  \centerline{\includegraphics[width=\linewidth]{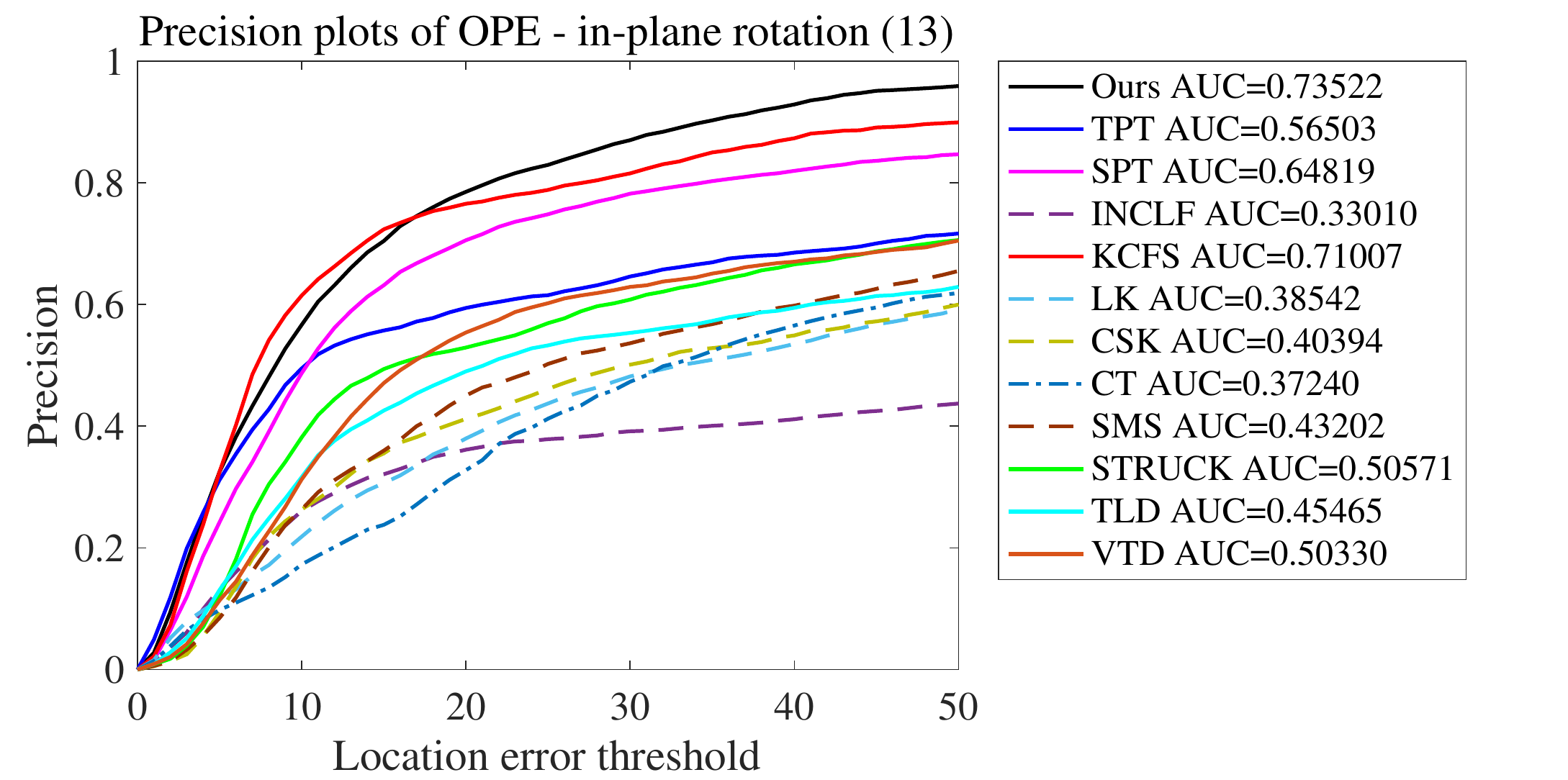}}  \vspace{4.pt}
  \end{minipage}
  \begin{minipage}[b]{0.3\linewidth}
  \centering
  \centerline{\includegraphics[width=\linewidth]{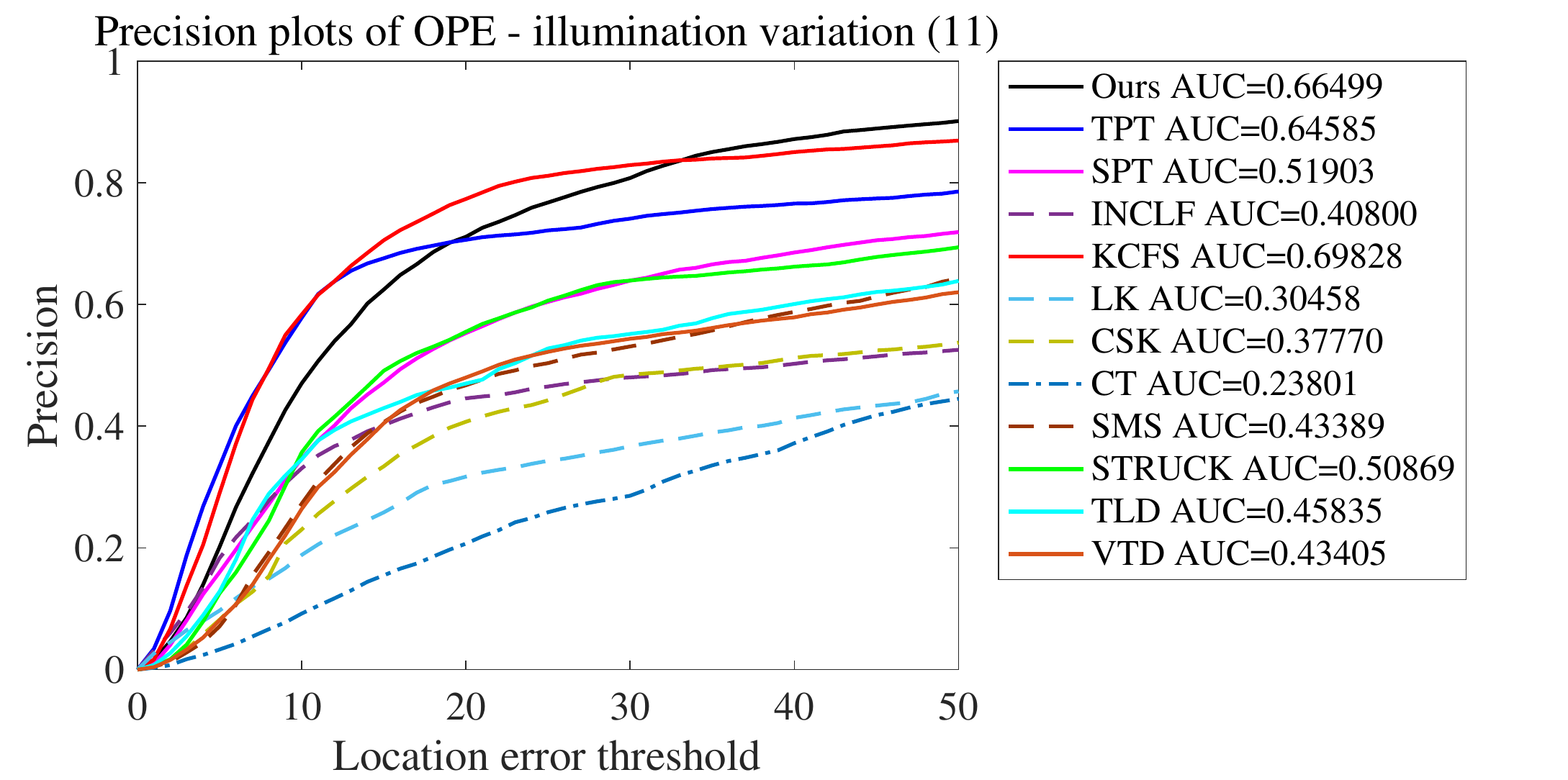}}\vspace{4.pt}
\end{minipage}
\begin{minipage}[b]{0.3\linewidth}
  \centering
  \centerline{\includegraphics[width=\linewidth]{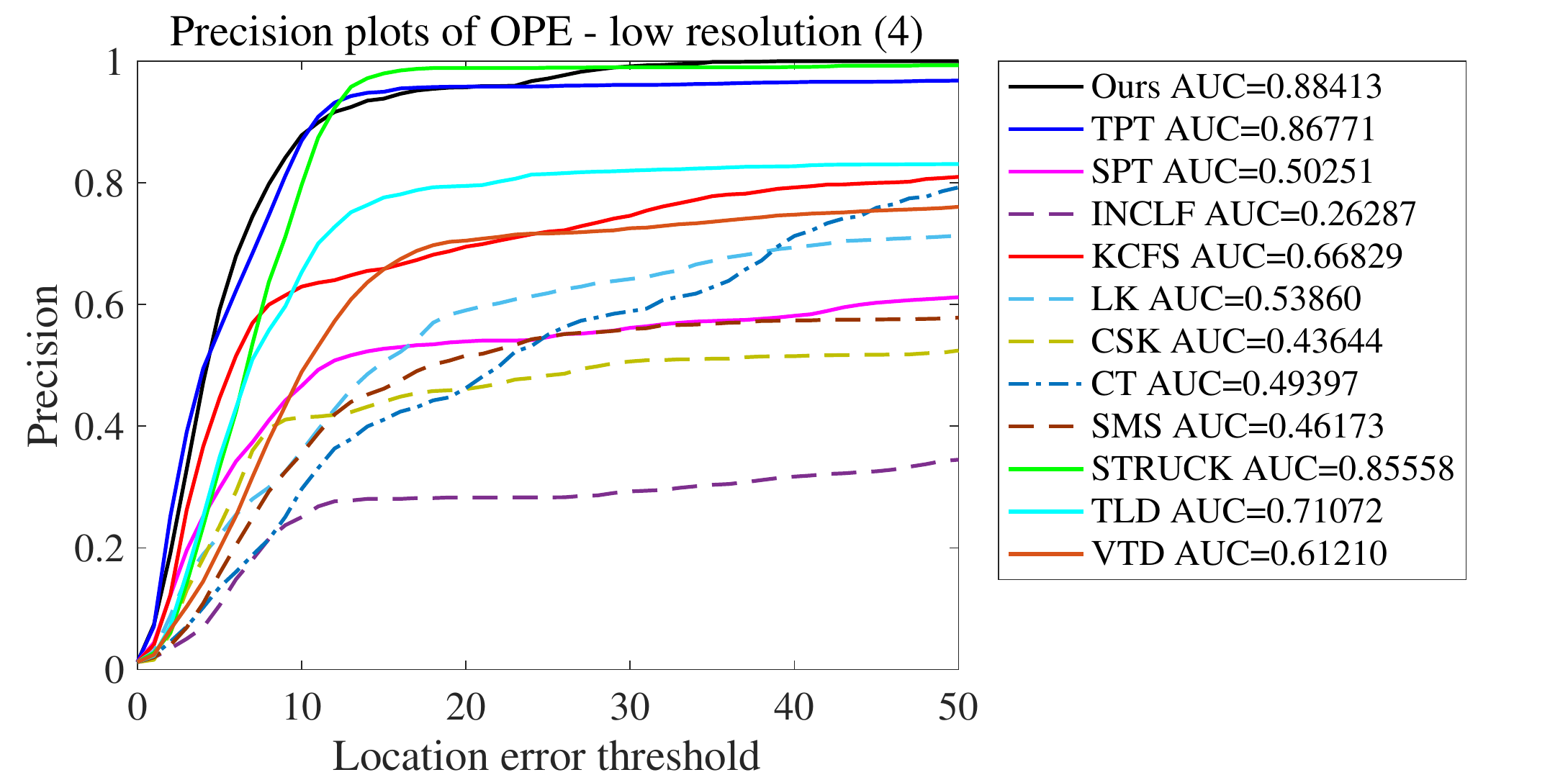}}\vspace{4.pt}
\end{minipage}
\begin{minipage}[b]{0.3\linewidth}
  \centering
  \centerline{\includegraphics[width=\linewidth]{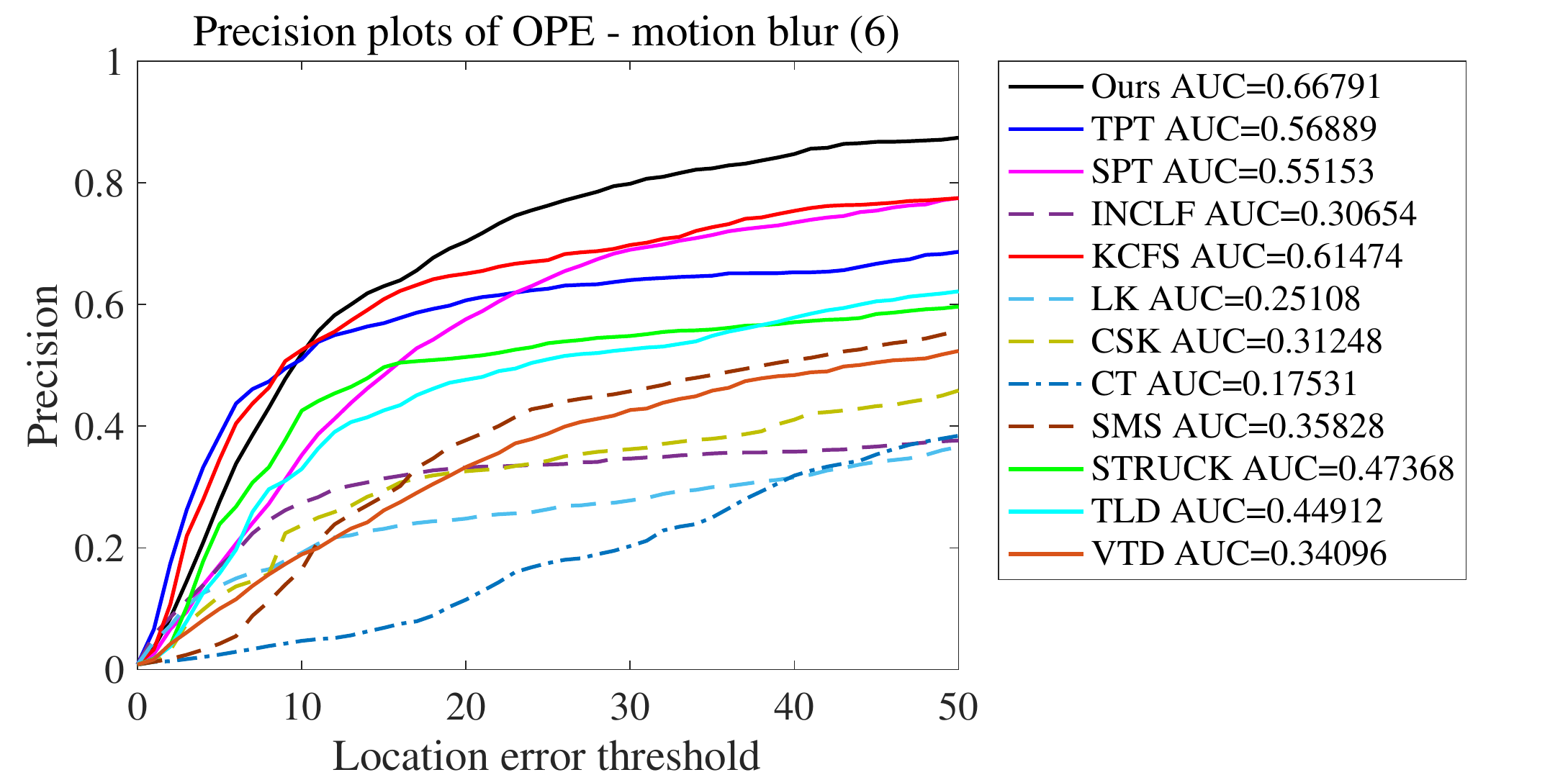}}\vspace{4.pt}
\end{minipage}
\begin{minipage}[b]{0.3\linewidth}
  \centering
  \centerline{\includegraphics[width=\linewidth]{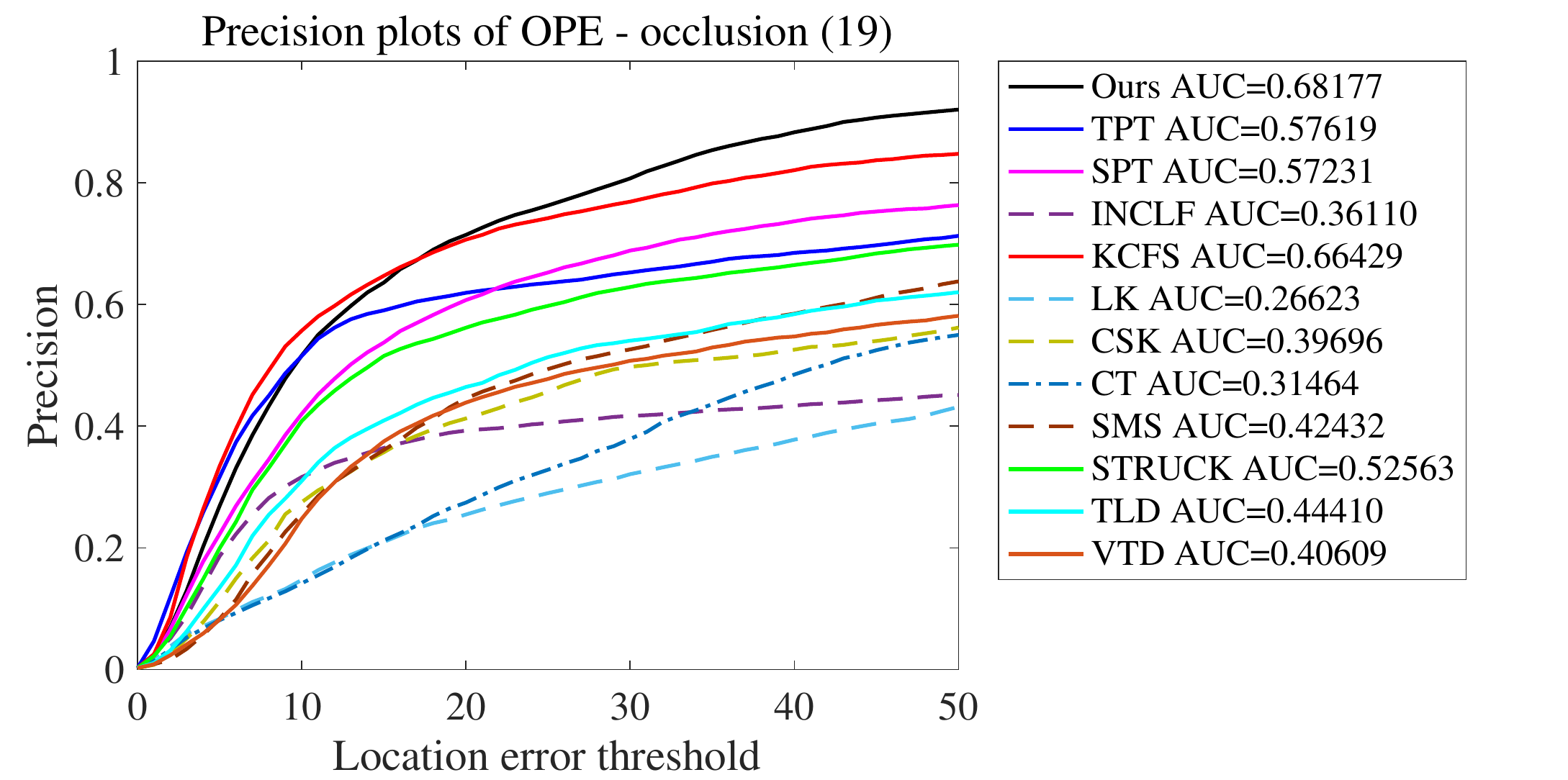}}  \vspace{4.pt}
\end{minipage}
  \begin{minipage}[b]{0.3\linewidth}
  \centering
  \centerline{\includegraphics[width=\linewidth]{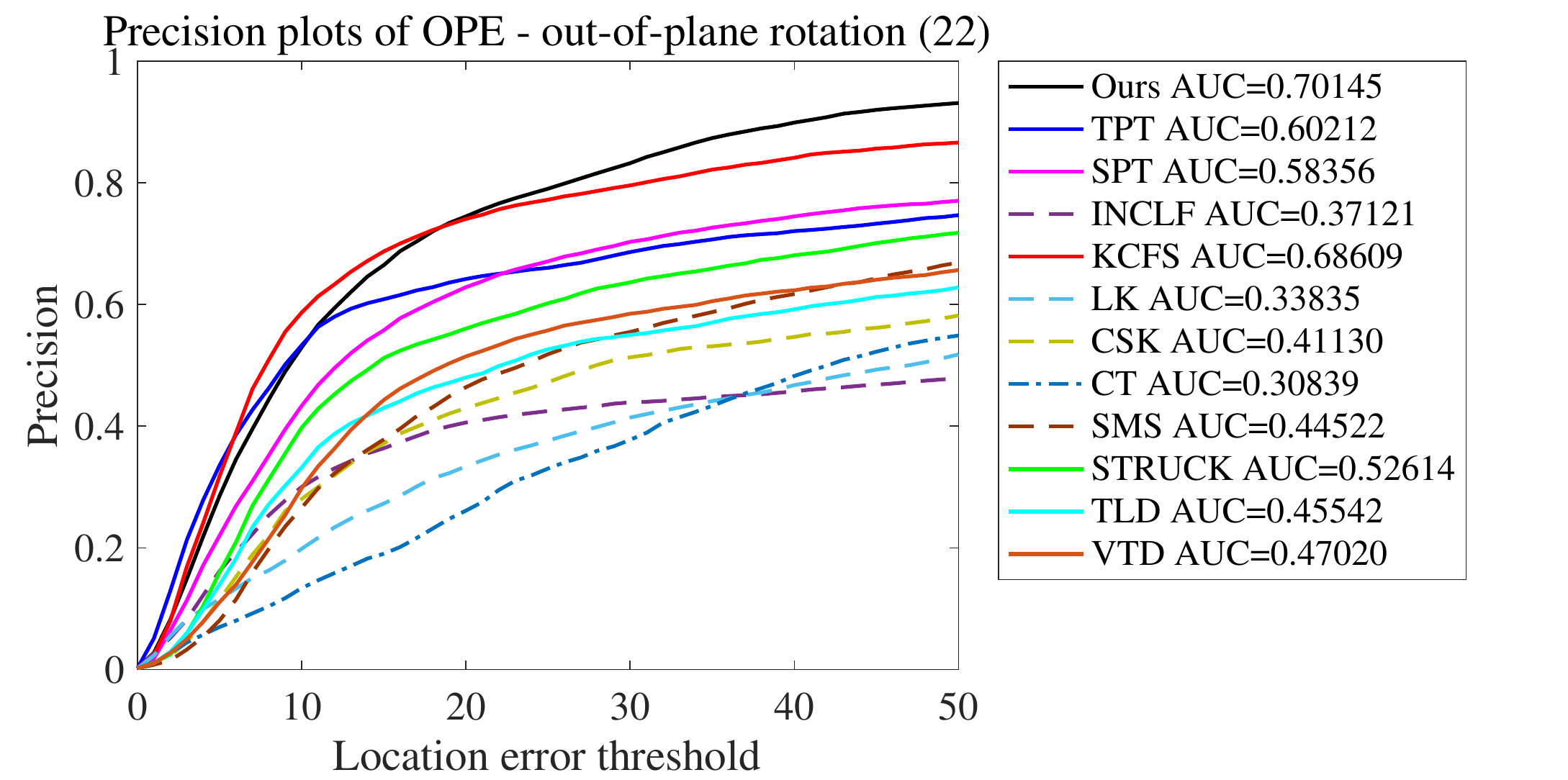}}\vspace{4.pt}
\end{minipage}
\begin{minipage}[b]{0.3\linewidth}
  \centering
  \centerline{\includegraphics[width=\linewidth]{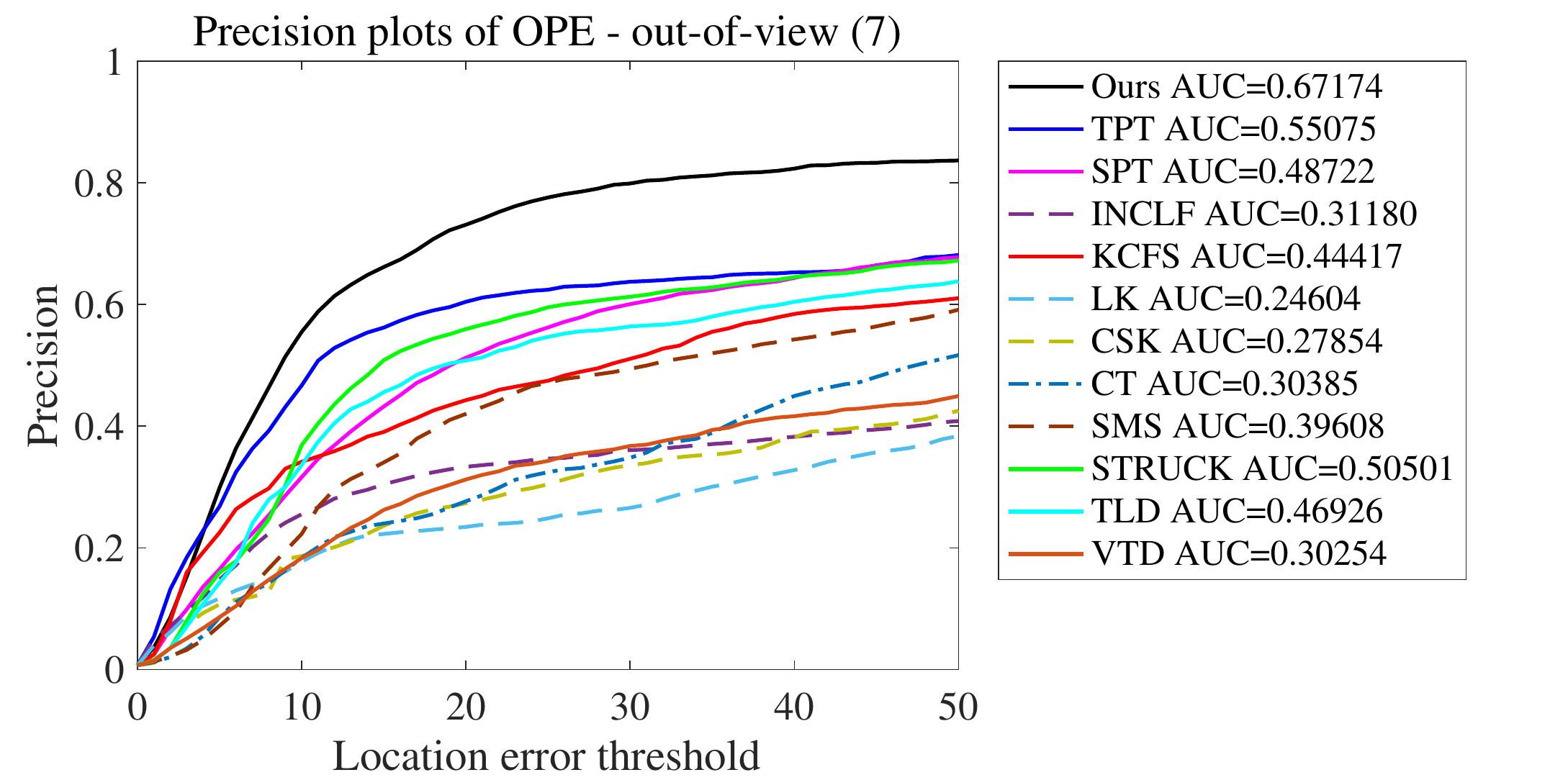}}\vspace{4.pt}
\end{minipage}
\begin{minipage}[b]{0.3\linewidth}
  \centering
  \centerline{\includegraphics[width=\linewidth]{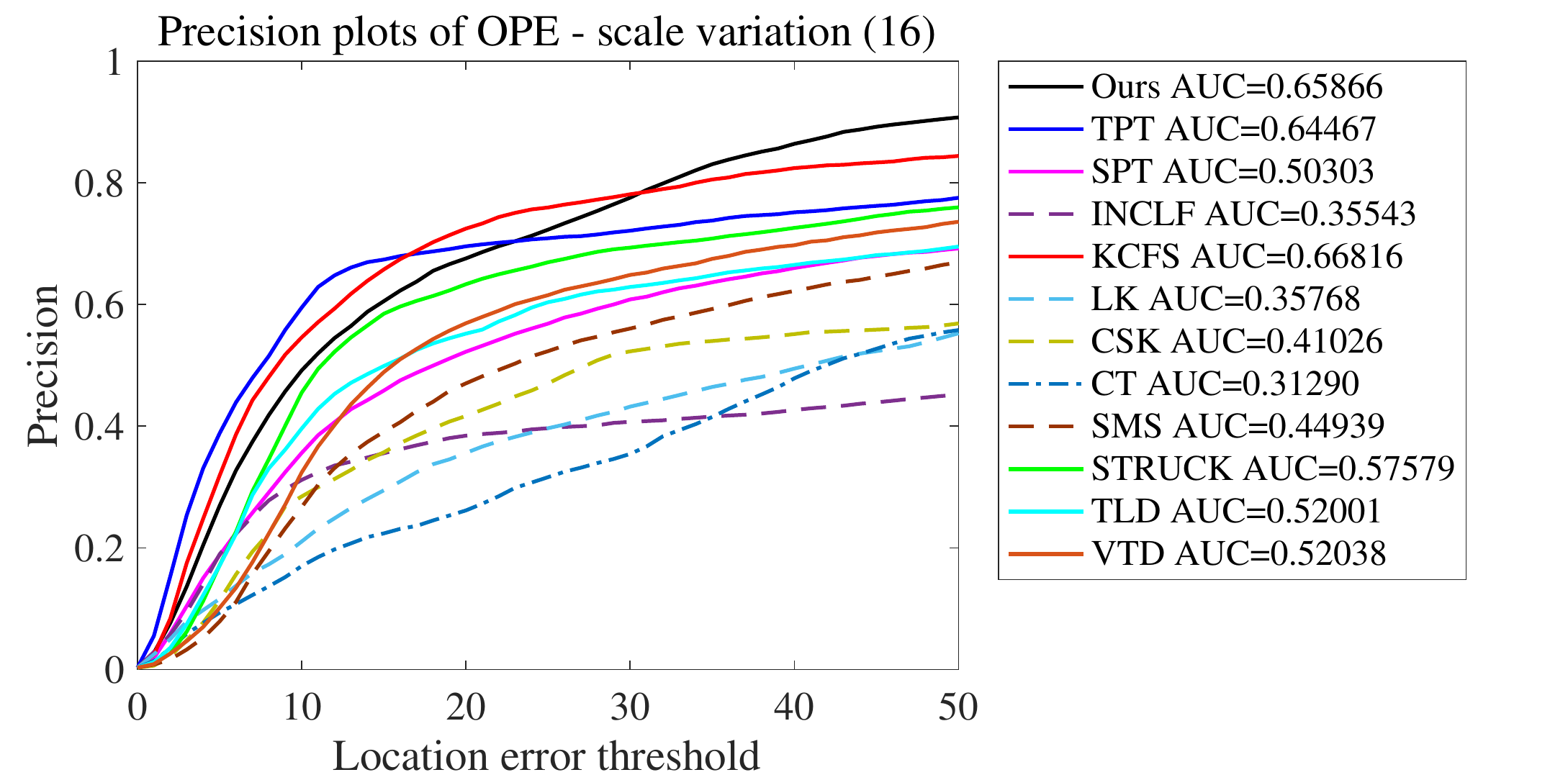}}\vspace{4.pt}
\end{minipage}
\caption{The precision plots of 12 trackers about 11 challenging attributes on the benchmark.}
\label{fig7}
\end{figure*}

Fig. \ref{fig4} shows some screen shots obtained by four top ranked trackers: the proposed tracker, SPT, TPT, and KCFS in the tracking of Bird1, Basketball, Skating2, and Bolt videos. In the Bird1 sequence, the target undergoes the tracking challenges of deformation, fast motion, and out-of-view. Our method can track more shots accurately than other three trackers. In the Basketball sequence, the target undergoes the tracking challenges of illumination variation, occlusion, deformation, out-of-plane rotation, and background clutters. SPT fails in tracking this video. Our method and KCFS achieve the comparable performance in this video and are better than TPT. In the sequence of Skating2, the target undergoes scale variation, occlusion, deformation, fast motion, and out-of-plane rotation. As shown in Fig. \ref{fig4}, our method can track the target more accurately. In the sequence of Bolt, the target undergoes occlusion, deformation, in-plane rotation, and out-of-plane rotation. TPT fails in tracking this video. To sum up, our method is more robust against multiple visual tracking challenges than other three trackers. Figs. \ref{fig2}-\ref{fig3} demonstrate the overall performance achieved by our method and state-of-the-art trackers. In Fig. \ref{fig2}, we can see that our method obtains highest success rate and outperforms the second one (KCFS) by 5.6\%. In Fig. \ref{fig3}, our method also obtains highest precision rate and outperforms the second one (KCFS) by 3.7\%. Figs. \ref{fig6}-\ref{fig7} illustrate attributed-based performance. From Fig. \ref{fig6}, we can see that our method achieves highest success rate in all challenging attributes except background clutters, illumination variation, low resolution, and sale variation. In terms of background clutters, the success rate of proposed tracker (58.6\%) is just inferior to KCFS (65.4\%). As to the illumination variation, the success rate achieved by our method (52.5\%) is almost the same compared to best score achieved by KCFS (53.0\%). In terms of low resolution and sale variation, our method also achieves the second best success rate. From Fig. \ref{fig7}, our method achieves highest precision rate in all challenging attributes except background clutters, illumination variation, and sale variation. In terms of background clutters, illumination variation, and sale variation, our method also achieves the second best precision rate. In conclusion, our method shows more robust against different visual tracking challenges compared to state-of-the-art trackers. As to the running time, compared with TPT, SPT, KCFS (three top ranked trackers): 2.3s per frame (Ours), 2.7s per frame (TPT), 0.8379s per frame (SPT), 0.0072s per frame (KCFS), our method is faster than TPT by 17.4\% but inferior to SPT and KCFS and fails to meet real-time requirement.

\section{Conclusion}
In this paper, we propose a visual tracking method which can fuse multiple midlevel cues obtained by superpixels to construct tensor-pooled sparse features. The discriminative model of our method has both good characteristics of the midlevel cues and sparse representation. In the validation, our method shows more robust against different visual tracking challenges than state-of-the-art methods. One main defect of our method is that it cannot meet the real-time requirement which is caused by large computation of IRTSA and tensor decomposition. We are studying using FPGA to implement an accelerator to speed up the tensor computation. In future, we will use the accelerator to implement a real time version of our method.

\section*{Acknowledgment}
This work was supported by the Hong Kong Research Grants Council
(Project C1007-15G) and City University of Hong Kong (Project 7005230).

\bibliographystyle{IEEEtran}
\bibliography{mybibfile}

\begin{thebibliography}{10}
\providecommand{\url}[1]{#1}
\csname url@samestyle\endcsname
\providecommand{\newblock}{\relax}
\providecommand{\bibinfo}[2]{#2}
\providecommand{\BIBentrySTDinterwordspacing}{\spaceskip=0pt\relax}
\providecommand{\BIBentryALTinterwordstretchfactor}{4}
\providecommand{\BIBentryALTinterwordspacing}{\spaceskip=\fontdimen2\font plus
\BIBentryALTinterwordstretchfactor\fontdimen3\font minus
  \fontdimen4\font\relax}
\providecommand{\BIBforeignlanguage}[2]{{%
\expandafter\ifx\csname l@#1\endcsname\relax
\typeout{** WARNING: IEEEtran.bst: No hyphenation pattern has been}%
\typeout{** loaded for the language `#1'. Using the pattern for}%
\typeout{** the default language instead.}%
\else
\language=\csname l@#1\endcsname
\fi
#2}}
\providecommand{\BIBdecl}{\relax}
\BIBdecl

\bibitem{SBD}
S.~Abdulhussain, A.~R. Ramli, B.~M. Mahmmod, M.~I. Saripan, S.~A.~R. Al-Haddad,
  and W.~A. Jassim, ``Shot boundary detection based on orthogonal polynomial,''
  \emph{Multimedia Tools and Applications}, pp. 1--22, 2019.

\bibitem{SPL1}
X.~{Zhan} and B.~{Ma}, ``Gaussian mixture model on tensor field for visual
  tracking,'' \emph{IEEE Signal Processing Letters}, vol.~19, no.~11, pp.
  733--736, Nov 2012.

\bibitem{TPT2}
L.~{Huang} and B.~{Ma}, ``Tensor pooling for online visual tracking,'' in
  \emph{2015 IEEE International Conference on Multimedia and Expo (ICME)}, June
  2015, pp. 1--6.

\bibitem{TPT}
B.~{Ma}, L.~{Huang}, J.~{Shen}, and L.~{Shao}, ``Discriminative tracking using
  tensor pooling,'' \emph{IEEE Transactions on Cybernetics}, vol.~46, no.~11,
  pp. 2411--2422, Nov 2016.

\bibitem{INCLF}
F.~{Liu}, T.~{Zhou}, C.~{Gong}, K.~{Fu}, L.~{Bai}, and J.~{Yang}, ``Inverse
  nonnegative local coordinate factorization for visual tracking,'' \emph{IEEE
  Transactions on Circuits and Systems for Video Technology}, vol.~28, no.~8,
  pp. 1752--1764, Aug 2018.

\bibitem{IBACF}
X.~{Sheng}, Y.~{Liu}, H.~{Liang}, F.~{Li}, and Y.~{Man}, ``Robust visual
  tracking via an improved background aware correlation filter,'' \emph{IEEE
  Access}, vol.~7, pp. 24\,877--24\,888, 2019.

\bibitem{BACF}
\BIBentryALTinterwordspacing
H.~Galoogahi, A.~Fagg, and S.~Lucey, ``Learning background-aware correlation
  filters for visual tracking,'' in \emph{2017 IEEE International Conference on
  Computer Vision (ICCV)}.\hskip 1em plus 0.5em minus 0.4em\relax Los Alamitos,
  CA, USA: IEEE Computer Society, oct 2017, pp. 1144--1152. [Online].
  Available: \url{https://doi.ieeecomputersociety.org/10.1109/ICCV.2017.129}
\BIBentrySTDinterwordspacing

\bibitem{KCFS}
W.~{Huang}, J.~{Gu}, X.~{Ma}, and Y.~{Li}, ``Correlation-filter based
  scale-adaptive visual tracking with hybrid-scheme sample learning,''
  \emph{IEEE Access}, vol.~6, pp. 125--137, 2018.

\bibitem{DLT}
N.~Wang and D.-Y. Yeung, ``Learning a deep compact image representation for
  visual tracking,'' in \emph{Advances in neural information processing
  systems}, 2013, pp. 809--817.

\bibitem{MDNet}
H.~Nam and B.~Han, ``Learning multi-domain convolutional neural networks for
  visual tracking,'' in \emph{The IEEE Conference on Computer Vision and
  Pattern Recognition (CVPR)}, June 2016.

\bibitem{HCF}
C.~Ma, J.~B. Huang, X.~K. Yang, and M.-H. Yang, ``Hierarchical convolutional
  features for visual tracking,'' in \emph{The IEEE International Conference on
  Computer Vision (ICCV)}, December 2015.

\bibitem{SPT}
S.~{Wang}, H.~{Lu}, F.~{Yang}, and M.~{Yang}, ``Superpixel tracking,'' in
  \emph{2011 IEEE International Conference on Computer Vision (ICCV)}, Nov
  2011, pp. 1323--1330.

\bibitem{SPT1}
F.~{Yang}, H.~{Lu}, and M.~{Yang}, ``Robust superpixel tracking,'' \emph{IEEE
  Transactions on Image Processing}, vol.~23, no.~4, pp. 1639--1651, April
  2014.

\bibitem{ISPT}
S.~D. {Lin} and D.~{Yang}, ``A visual tracking method using superpixel,'' in
  \emph{2017 International Conference on Machine Learning and Cybernetics
  (ICMLC)}, vol.~2, July 2017, pp. 552--557.

\bibitem{CST}
L.~{Wang}, H.~{Lu}, and M.~{Yang}, ``Constrained superpixel tracking,''
  \emph{IEEE Transactions on Cybernetics}, vol.~48, no.~3, pp. 1030--1041,
  March 2018.

\bibitem{DSP}
Y.~{Yuan}, J.~{Fang}, and Q.~{Wang}, ``Robust superpixel tracking via depth
  fusion,'' \emph{IEEE Transactions on Circuits and Systems for Video
  Technology}, vol.~24, no.~1, pp. 15--26, Jan 2014.

\bibitem{SLIC}
R.~Achanta, A.~Shaji, K.~Smith, A.~Lucchi, P.~Fua, and S.~Süsstrunk, ``{SLIC}
  superpixels compared to state-of-the-art superpixel methods,'' \emph{IEEE
  Transactions on Pattern Analysis and Machine Intelligence}, vol.~34, no.~11,
  pp. 2274--2282, Nov 2012.

\bibitem{SNIC}
R.~{Achanta} and S.~{Süsstrunk}, ``Superpixels and polygons using simple
  non-iterative clustering,'' in \emph{2017 IEEE Computer Vision and Pattern
  Recognition (CVPR)}, July 2017, pp. 4895--4904.

\bibitem{FSLIC}
C.~Wu, L.~Zhang, H.~Zhang, and H.~Yan, ``Fuzzy {SLIC}: Fuzzy simple linear
  iterative clustering,'' \emph{arXiv preprint arXiv:1812.10932}, 2018.

\bibitem{ICIP1}
C.~{Wu}, L.~{Zhang}, H.~{Zhang}, and H.~{Yan}, ``Improved superpixel-based fast
  fuzzy {C}-means clustering for image segmentation,'' in \emph{2019 IEEE
  International Conference on Image Processing (ICIP)}, Sep. 2019, pp.
  1455--1459.

\bibitem{JFast}
S.~Jia and C.~Zhang, ``Fast and robust image segmentation using an superpixel
  based {FCM} algorithm,'' in \emph{2014 IEEE International Conference on Image
  Processing (ICIP)}, Oct 2014, pp. 947--951.

\bibitem{FuzzyS}
Y.~Guo, L.~Jiao, S.~Wang, S.~Wang, F.~Liu, and W.~Hua, ``Fuzzy superpixels for
  polarimetric {SAR} images classification,'' \emph{IEEE Transactions on Fuzzy
  Systems}, vol.~26, no.~5, pp. 2846--2860, Oct 2018.

\bibitem{Harmony}
X.~Boix, J.~M. Gonfaus, J.~Weijer, A.~D. Bagdanov, and J.~Serrat, ``Harmony
  potentials,'' \emph{International Journal of Computer Vision (IJCV)},
  vol.~96, no.~1, pp. 83--102, 2012.

\bibitem{SEEDS}
M.~V.~D. Bergh, X.~Boix, G.~Roig, B.~D. Capitani, and L.~V. Gool, ``{SEEDS}:
  superpixels extracted via energy-driven sampling,'' in \emph{European
  Conference on Computer Vision (ECCV)}.\hskip 1em plus 0.5em minus 0.4em\relax
  Springer, 2012, pp. 13--26.

\bibitem{SPARSE}
C.~{Huang} and B.~{Jiang}, ``Occlusion handling of visual tracking by fusing
  multiple visual clues,'' in \emph{2017 IEEE International Conference on
  Systems, Man, and Cybernetics (SMC)}, Oct 2017, pp. 836--839.

\bibitem{Bench}
Y.~{Wu}, J.~{Lim}, and M.~{Yang}, ``Object tracking benchmark,'' \emph{IEEE
  Transactions on Pattern Analysis and Machine Intelligence}, vol.~37, no.~9,
  pp. 1834--1848, Sep. 2015.

\bibitem{IRTSA1}
X.~{Li}, W.~{Hu}, Z.~{Zhang}, X.~{Zhang}, and G.~{Luo}, ``Robust visual
  tracking based on incremental tensor subspace learning,'' in \emph{2007 IEEE
  International Conference on Computer Vision (ICCV)}, Oct 2007, pp. 1--8.

\bibitem{PF}
M.~{Isard} and A.~{Blake}, ``Contour tracking by stochastic propagation of
  conditional density,'' in \emph{European Conference on Computer
  Vision}.\hskip 1em plus 0.5em minus 0.4em\relax Springer, 1996, pp. 343--356.

\bibitem{LK}
M.~Lucena, J.~Fuertes, N.~de~la Blanca, and A.~Garrido, ``Using optical flow as
  evidence for probabilistic tracking,'' in \emph{Scandinavian Conference on
  Image Analysis}.\hskip 1em plus 0.5em minus 0.4em\relax Springer, 2003, pp.
  1044--1049.

\bibitem{STRUCK}
S.~{Hare}, A.~{Saffari}, and P.~H.~S. {Torr}, ``Struck: Structured output
  tracking with kernels,'' in \emph{2011 IEEE International Conference on
  Computer Vision (ICCV)}, Nov 2011, pp. 263--270.

\bibitem{TLD}
Z.~{Kalal}, J.~{Matas}, and K.~{Mikolajczyk}, ``P-{N} learning: Bootstrapping
  binary classifiers by structural constraints,'' in \emph{2010 IEEE Computer
  Vision and Pattern Recognition (CVPR)}, June 2010, pp. 49--56.

\bibitem{VTD}
J.~{Kwon} and K.~M. {Lee}, ``Visual tracking decomposition,'' in \emph{2010
  IEEE Computer Vision and Pattern Recognition (CVPR)}, June 2010, pp.
  1269--1276.

\bibitem{CSK}
F.~{Henriques}, R.~{Caseiro}, P.~{Martins}, and J.~{Batista}, ``Exploiting the
  circulant structure of tracking-by-detection with kernels,'' in
  \emph{European Conference on Computer Vision (ECCV)}.\hskip 1em plus 0.5em
  minus 0.4em\relax Springer, 2012, pp. 702--715.

\bibitem{SMS}
R.~T. {Collins}, ``Mean-shift blob tracking through scale space,'' in
  \emph{2003 IEEE Computer Vision and Pattern Recognition (CVPR)}, vol.~2, June
  2003, pp. II--234.

\bibitem{CT}
K.~{Zhang}, L.~{Zhang}, and M.-H. {Yang}, ``Real-time compressive tracking,''
  in \emph{European Conference on Computer Vision (ECCV)}.\hskip 1em plus 0.5em
  minus 0.4em\relax Springer, 2012, pp. 864--877.

\end{thebibliography}

\end{document}